# Sensitivity of Quantitative Susceptibility Mapping in Clinical Brain Research

Fahad Salman[1,5] (Fahadsal@buffalo.edu), Abhisri Ramesh[1] (abhisri.ramesh@gwmail.gwu.edu), Thomas Jochmann[1,2] (thomas@jochmann.info), Mirjam Prayer[1] (mirjam.prayer@tu-dortmund.de), Ademola Adegbemigun[1] (ademolaa@buffalo.edu), Jack A. Reeves[1] (jackreev@buffalo.edu), Gregory E. Wilding[3] (gwilding@buffalo.edu), Junghun Cho[1,5] (jcho34@buffalo.edu), Dejan Jakimovski[1] (djakimovski@bnac.net), Niels Bergsland[1] (npbergsland@bnac.net), Michael G. Dwyer[1,4] (mgdwyer@bnac.net), Robert Zivadinov[1,4] (rzivadinov@bnac.net), and Ferdinand Schweser[*,1,4] (schweser@buffalo.edu)

[1]Buffalo Neuroimaging Analysis Center, Department of Neurology at the Jacobs School of Medicine and Biomedical Sciences, University at Buffalo, The State University of New York, Buffalo, NY, United States; [2]Department of Computer Science and Automation, Technische Universität Ilmenau, Ilmenau, Germany; [3]Department of Biostatistics, School of Public Health and Health Professions, State University of New York at Buffalo, Buffalo, NY, USA; [4]Center for Biomedical Imaging, Clinical and Translational Science Institute, University at Buffalo, The State University of New York, Buffalo, NY, United States; [5]Department of Biomedical Engineering, University at Buffalo, The State University of New York, Buffalo, NY, United States

* corresponding author

**Word count:** 8,500

**Acknowledgments**

We are grateful to Pascal Spincemaille (Cornell-Weill Medical College, USA) for his valuable advice on the PDF implementation for single-echo data, Hongfu Sun (University of Calgary, Canada) for regularization parameter suggestion for the RESHARP algorithm, Shuai Huang (Emory University, Canada) for AMP-PE implementation and optimization, and Carlos Milovic (Pontificia Universidad Católica de Valparaíso, Chile) for his support in implementing the WH-FANSI, L1-QSM, and HD-QSM methods and for parameter optimization. We thank Steffen Bollmann (University of Queensland, Australia) for providing the DeepQSM method and Jongho Lee (Seoul National University, South Korea) for the QSMnet+ implementation. Research reported in this publication was partially supported by the National Institute of Neurological Disorders And Stroke of the National Institutes of Health under Award Number R01NS114227, the National Center for Advancing Translational Sciences of the National Institutes of Health under Award Number UL1TR001412, the German Federal Ministry of Education and Research (BMBF) grant AVATAR (16KISA024, funded by the European Union - NextGenerationEU), the German Academic Exchange Service (DAAD PPP 57599925), and an ISMRM Research Exchange Grant awarded to T.J. The content is solely the responsibility of the authors and does not necessarily represent the official views of the National Institutes of Health.

**Abstract**

**Background**: Quantitative susceptibility mapping (QSM) of the brain is an advanced MRI technique for assessing tissue characteristics based on magnetic susceptibility, which varies with the composition of the tissue, such as iron, calcium, and myelin levels. QSM consists of multiple processing steps, with various choices for each step. Despite its increasing application in detecting and monitoring neurodegenerative diseases, the impact of algorithmic choices in QSM's workflow on clinical outcomes has not been thoroughly quantified.

**Objective**: This study aimed to evaluate how choices in background field removal (BFR), dipole inversion algorithms, and anatomical referencing impact the sensitivity and reproducibility error of QSM in detecting group-level and longitudinal changes in deep gray matter susceptibility in a clinical setting.

**Methods**: We compared 378 different QSM pipelines using a 10-year follow-up dataset of healthy adults scanned at a 3T. We analyzed the ability of these pipelines to detect known aging-related susceptibility changes in the DGM over time, linked to altered iron patterns established in earlier foundational studies.

**Results**: Our results demonstrated high variability in the sensitivity of QSM pipelines to detect susceptibility changes. The study highlighted that while most pipelines could detect changes reliably, the choice of BFR algorithm and the referencing strategy substantially influenced the outcome reproducibility error and sensitivity. Notably, pipelines using RESHARP with AMP-PE, HEIDI or LSQR inversion showed the highest overall sensitivity.

**Conclusions**: The findings underscore the critical influence of algorithmic choices in QSM processing on the accuracy and reliability of detecting physiological changes in the brain. This has profound implications for clinical research and trials where QSM is used as a biomarker for disease progression, highlighting that careful consideration should be given to pipeline configuration to optimize clinical outcomes. Our study also indicates the need for standardized protocols in QSM to enhance comparability across studies.

**Key Points**

- We analyzed the sensitivity of different quantitative susceptibility mapping (QSM) pipelines ($N$=378; background field removal + dipole inversion + reference region) towards the detection of known aging-related susceptibility changes in the deep gray matter over a period of 10 years. Our results demonstrated high variability in the sensitivity of QSM pipelines to detect susceptibility changes.
- The study highlighted that while most pipelines could detect changes reliably, the choice of BFR algorithm and the referencing strategy substantially influenced the outcome reproducibility error and sensitivity.
- In clinical research and trials utilizing QSM as a disease progression biomarker, it is important to consider the performance of the entire QSM pipeline rather than the individual components in isolation.

**Declaration of Conflicting Interests**

R.Z. has received personal compensation from Bristol Myers Squibb, EMD Serono, Sanofi, Mapi Pharma, Sana Biotechnologies and Filterlex for speaking and consultant fees. He received financial support for research activities from Bristol Myers Squibb, EMD Serono, Mapi Pharma and Protembis and Filterlex.

M.D. received personal compensation from Bristol Myers Squibb, Novartis, EMD Serono and Keystone Heart, and financial support for research activities from Bristol Myers Squibb, Novartis, Mapi Pharma, Keystone Heart, Protembis, and V-WAVE Medical.

**Ethics Approval**

The local Ethical Standards Committee approved the human experiments, and a written informed consent form was obtained.

**Data Availability**

The imaging data supporting the findings of this study are available from the corresponding author upon reasonable request and subject to institutional review board approval. These data are not publicly accessible due to privacy and ethical considerations.

**Author Contributions (CRediT)**

Ademola Adegbemigun - Methodology

Niels Bergsland - Conceptualization, Data curation, Resources, Software, Writing – review & editing

Junghun Cho - Formal Analysis, Investigation, Writing – review & editing

Michael G. Dwyer - Conceptualization, Data curation, Resources, Software

Dejan Jakimovski – Data curation

Thomas Jochmann - Methodology, Software, Writing – review & editing

Mirjam Prayer - Methodology, Methodology

Abhisri Ramesh - Methodology, Software

Jack A. Reeves - Formal Analysis, Investigation, Writing – review & editing

Fahad Salman - Conceptualization, Data curation, Formal Analysis, Investigation, Methodology, Project administration, Software, Validation, Visualization, Writing – original draft, Writing – review & editing

Ferdinand Schweser - Conceptualization, Funding acquisition, Investigation, Methodology, Project administration, Resources, Software, Supervision, Validation, Writing – original draft, Writing – review & editing

Gregory E. Wilding -  Methodology, Supervision

Robert Zivadinov - Conceptualization, Data curation, Resources, Writing – review & editing

# Introduction

Quantitative susceptibility mapping (QSM) is a magnetic resonance imaging (MRI) technique that calculates tissue magnetic susceptibility from gradient-echo phase images.[1,2] The technique can be used to assess tissue iron (paramagnetic)[3], myelin[4], and calcium (both diamagnetic).[5] The use of QSM in clinical research is growing rapidly[79,80], not only in the brain but also outside the brain (for e.g., liver and kidney).[6-9]

In the deep gray matter (DGM) of the brain, magnetic susceptibility correlates strongly with tissue iron concentrations.[3] Several neurodegenerative diseases, including multiple sclerosis[76-78], Parkinson's disease[10-12], and Alzheimer's disease [13,14], have been found to show altered magnetic susceptibility in DGM regions. QSM may serve as a component in imaging biomarkers, such as iron changes over-time and between group differences, for monitoring disease progression in these and other conditions, which would be highly valuable as an outcome measure in clinical trials.

QSM involves several integral processing steps, with the final three being (i) background field removal (BFR), (ii) the solution of an ill-posed dipole inversion problem, and (iii) referencing of susceptibility values to an internal reference point.[2] Background fields are the field contributions from sources located outside the region of interest (e.g., brain), which are removed using a BFR algorithm. Following BFR, dipole inversion is performed to derive a tissue magnetic susceptibility map ($\chi$) from the corrected field map. For both the BFR and the dipole inversion. Numerous algorithms have been proposed for BFR and dipole inversion over the past decade. The QSM community has benchmarked inversion algorithms in two challenges since 2016.[17,18] Both challenges used a single-subject gold standard susceptibility map to assess the fidelity of the susceptibility maps computed with different algorithms.[19] The challenges provided numerous insights into the state of the field and methodological challenges, such as substantial differences in susceptibility map appearance between algorithms. Since the investigations only used one subject susceptibility map, the *practical (*or clinical) significance of these benchmarking efforts remains limited. Specifically, it remains unknown if the algorithms with the highest performance in the QSM challenges also offer the best sensitivity toward detecting group differences (cross-sectionally) or changes in susceptibility over time (longitudinally) in a clinical research setting. Furthermore, it remained unknown how the BFR affects sensitivity.

Following the BFR and dipole inversion steps, susceptibility values need to be referenced to an internal reference region[20] before susceptibility findings can be reported.[21] While there is consensus that the reference region choice can impact statistical power and introduce bias[21], the optimal anatomical region for referencing remains a matter of ongoing debate[21] and the effect of referencing on group-level or longitudinal study outcomes is poorly understood.

The overarching aim of this study was to understand whether clinical study outcomes are affected by the specific choices of BFR algorithm, dipole inversion algorithm, and referencing region, referred to as QSM "pipeline" throughout this work. The goal was to identify methods that detect group-level susceptibility differences with the highest sensitivity in a typical clinical research setting — a pursuit largely unexplored until now, with previous emphasis primarily placed on the reproducibility and the accuracy of the dipole inversion step.[63-66]

# Theory

The effect size $(d)$ is the relevant statistical quantity that describes the strengths of an observation in a cohort-based study. In any real-world QSM study, the observed $d_R$ of a true susceptibility difference $\Delta\chi_R$ in a region of interest (ROI), $R$, between two sample distributions can be defined as

$$d_R = \frac{\Delta\chi_R}{\sigma_{overall,R}}, \tag{1}$$

where $\sigma_{overall,R}$ is the average of intra-group standard deviations of observed susceptibility values, which follows from the propagation of uncertainty as

$$\sigma^2_{overall,R} = \sigma^2_{true,R} + \sigma^2_{reference} + \sigma^2_{measure,R}, \tag{2}$$

where $\sigma_{true,R}$ is the standard deviation of the true tissue susceptibility in the ROI, which is given by the biological differences between subjects, $\sigma_{reference}$ is the variation due to natural biological variability of the true susceptibility in the reference region, and $\sigma_{measure,R}$ is the variation caused by the measurement process. The quantity $\sigma_{measure,R}$ can be decomposed into variations due to reconstruction artifacts within the ROI, $\sigma_{artifacts,R}$ and within the reference region $\sigma_{artifacts,reference}$:

$$\sigma^2_{measure,R} = \sigma^2_{artifacts,R} + \sigma^2_{artifacts,reference}. \tag{3}$$

In clinical studies, it is desirable to use algorithms that detect susceptibility differences with a high absolute effect size, $|d|$. To achieve a high absolute effect size, the absolute susceptibility difference in Eq. (1) must be high (determined by the biology) and the variation caused by the measurement process, $\sigma_{\text{measure},R}$, and natural variation in the reference region, $\sigma^2_{\text{reference}}$, must be low.

In an ideal world, the variation due to the measurement process would be zero, $\sigma^2_{\text{measure},R} = 0$, and the effect size would be determined entirely by biological variability, $\sigma^2_{overall,R} \approx \sigma^2_{true,R} + \sigma^2_{\text{reference}}$. However, in the real world, the presence of artifacts is unavoidable, e.g., due to motion during the data acquisition and field inhomogeneities.[2]

# Materials & Methods

## Study Design

Real-world evaluations of QSM have been limited by the lack of a ground truth for in vivo susceptibility quantification.[17] In the present study, we relied on the premise that aging-related changes in DGM susceptibility are driven by well-established age-related changes in iron concentration.[81] In their 1958 landmark paper, Hallgren and Sourander[22] (H&S) showed that histochemically determined iron concentrations followed an exponential saturation trajectory with age in most brain regions. The rate of age-related iron concentration changes varied between regions, with putamen and caudate increasing for the longest time throughout adulthood (see p. 48 therein). In the thalamus, the authors found that iron concentrations decreased starting in the fourth decade of life (Fig. 8 therein).[22] The iron trajectories reported by H&S have been replicated successfully using QSM in a cross-sectional study.[81]

Based on this evidence, we presumed that properly functioning QSM pipelines should detect increasing susceptibility *over time* in the globus pallidus (GP), putamen, and caudate of healthy adults ($d > 0$) and declining susceptibility in the thalamus ($d < 0$ for age >30 years) if all other systematically contributing biological factors to tissue susceptibility can be assumed to be negligible. Myelin, which is the major contributor to susceptibility differences in the brain besides iron[23], appears to undergo its most rapid increase in the first three decades, followed by relative stability and a slow decline thereafter, with myelination in the gray matter (GM) remaining relatively stable (from $3^{rd}$ to $6^{th}$ decade; refer to Fig. 5 and 6 of the original publication, top left panel).[24]

We used the effect size of aging-related changes in susceptibility as a key performance metric to compare algorithmic combinations. In this context, the effect size represented the sensitivity toward the detection of aging-related effects and may be considered as a surrogate metric of general sensitivity of QSM toward cross-sectional group differences or over-time changes in the clinical research setting.

**Study Participants**

This prospective study enrolled *N*=25 subjects without neurological disease. Eligible study participants were selected from an institutional database of healthy controls that had participated in previous IRB-approved studies at our institution and provided their informed consent. The database was filtered concerning the availability of raw k-space data for a specific gradient-echo imaging sequence at a specific 3T MRI instrument and sorted for the first exam date on which the sequence had been applied. To minimize confounding effects from myelin, we restricted enrollment to subjects older than 35 years. Identified subjects were called in for a follow-up exam in the order of the date of the first available exam, starting with the subjects that had the earliest exam to maximize follow-up time in the resulting cohort.

We recruited another $N=5$ subjects without neurological disease for a scan-rescan experiment.

All study procedures were approved by the Institutional Review Board of the University at Buffalo and all participants ($N=30$) provided written informed consent in accordance with the Declaration of Helsinki.

**Imaging**

We used the same MRI scanner (3T GE Signa Excite HDx 23.0 scanner; General Electric, Milwaukee, WI, USA) for the follow-up experiment that was also used for the baseline exam. The scanner used an eight-channel head-and-neck coil. There were no major hard- or software upgrades to the MRI system between the baseline and follow-up exams. The data for QSM was collected at baseline using a 3D single-echo spoiled gradient recalled echo (GRE) sequence with first-order flow compensation in the read and slice directions, a matrix of 512x192x68 and a nominal resolution of 0.5x0.5x2mm$^3$ (FOV=256x192x128mm$^3$), flip angle=12°, TE/TR=22ms/40ms, bandwidth=13.89 kHz. We used identical pulse sequence parameters for the follow-up exam. Each channel's raw k-space data was retained for offline image reconstruction. Additionally, the following sequence was acquired for brain segmentation (see below; also identical between baseline and follow-up): Fast spoiled gradient-echo pulse sequence with inversion recovery (IR-FSPGR) for T1w imaging using the following parameters: TE/TI/TR=2.8ms/900ms/5.9ms, matrix=256×192×128 matrix, nominal resolution of 1×1×1 mm$^3$ (FOV=256×192×192mm$^3$), and flip angle=10°.

Scan-rescan subjects received the same GRE sequence four times, interleaved with removal of the subject from the magnet, full repositioning and recalibration of the MRI system. We prescribed each scan axial-oblique with slight intentional variations of the prescription between repetitions to ensure that variations due to image slab prescription would be captured in the data.

**Scientific Rigor – QSM processing**

To ensure optimal method implementation, all processing for QSM relied on code provided by the original developers, unless explicitly stated. To ensure high scientific rigor and avoid unintentional bias towards our in-house developed algorithms, we performed all processing fully automated and reproducible using containerized computing on a high-performance cluster with 4 CPUs (Intel Xeon Gold 6330) and 20 GB of memory allocated to each job. We used an in-house developed job scheduler (Simple Linux Utility for Resource Management – SLURM) for multi-step pipelines with dependencies (pi4s; https://gitlab.com/R01NS114227/pi4s). All QSM related processes (reconstructions and pre-processing for deep learning methods) were executed in a custom Singularity container generated with Neurodocker (https://www.repronim.org/neurodocker/), ensuring that the computational environment was identical for all methods. The container included Matlab (version 2018b; The MathWorks, Natick, MA), Freesurfer (v6.0.0), FSL (v5.0.8), and ANTs (v2.0.0). For the deep learning methods, we use a containerized version of Python (with Tensorflow). Computer code used for the processing has been made available at https://doi.org/10.5281/zenodo.11077423.[109] (note to reviewers: the archive will be set public upon acceptance of the paper).

**Processing**

*Phase image reconstruction*

The GRE images for QSM were reconstructed offline from the raw k-space data on a 512x512x64 spatial matrix using sum-of-squares for magnitude images and scalar phase matching for phase images.[25] The k-space data were zero-padded in the phase-encoding direction before the processing to achieve an isotropic in-plane resolution, in line with the manufacturer-based image reconstruction. Distortions due to imaging gradient non-linearity were compensated for as previously described.[26] Subsequently, phase images were unwrapped utilizing a best-path algorithm.[27]

*Included algorithms and reference regions*

<u>BFR algorithms:</u> We included the following widely used and publicly available BFR methods: Improved HARmonic (background) PhasE REmovaL using the LAplacian (iHARPERELLA)[32], Laplacian Boundary Value (LBV)[28], Projection onto Dipole Fields (PDF)[33], Regularization Enabled SHARP (RESHARP)[31], Sophisticated Harmonic Artifact Reduction for Phase (SHARP)[29], and variable-radius SHARP (V-SHARP).[29,30]

<u>Inversion algorithms</u>: The selection of inversion algorithms was guided by the results of the QSM Reconstruction Challenge 2.0.[18] We included the top-ranking inversion algorithms (*N*=5) from the normalized root mean square error (NRMSE) category, with one (MEDI)[44] algorithm from stage 1 and four (FANSI[39], HD-QSM[40], L1-QSM[42], Weak Harmonic QSM [WH-FANSI][50]) from stage 2 results of the report.[18] Additionally, we considered algorithms (*N*=14) that were publicly available, developed in our lab, or shared with us directly for inclusion (DeepQSM[36], DeepQSM with in-house generated training data[37] [DeepQSM*], DirTIK[38], HEIDI[41], iLSQR[51,] LSQR[41],

MEDI+0[45], MEDI+0 without CSF-regularization[45] [MEDI+0*], MATV[43], SDI[47], STAR[48], TKD[47,49], IterTIK[38], iSWIM[52]).

Furthermore, we invited authors of methods not yet included in the study to provide their algorithms for inclusion in the evaluation during presentation of the design and preliminary results of this study at the 2023 Annual Meeting of the ISMRM[82]. The methods AMP-PE[35] and QSMnet+[46] were included based on this request.

*Included reference regions*

Reference regions were chosen based on frequent use in the literature, past evaluations[20,66], as well as inclusion into the recent QSM consensus statements[21]: whole brain (WB; Supplemental Materials [Supp. Fig. 1])[76,104,105], white matter (WM)[3,101-103], and cerebrospinal fluid (CSF).[4,12,51,100]

In total, we compared $N$=378 (6 BFRs × 21 inversions × 3 reference regions) pipelines. We used default parameters for all algorithms or parameters suggested by the original authors in their original publications. If default or publication-guided parameters resulted in apparent artifacts or results that differed qualitatively from those presented in the original publications, we contacted the corresponding author of the original publication and asked for assistance with the implementation of algorithm.

*Background field removal*

The initial brain mask (*mask 1*) was generated by applying a whole-brain segmentation tool – FSL-brain extraction tool (BET)[53] to the magnitude images. This mask removed air, skull, and other tissues while preserving cortical areas. Following this, a mask of reliable phase values

(*mask 2*) was generated by thresholding the two-pixel finite difference of the unwrapped phase at an empirically determined value of 2.6. Subsequently, *masks 1* and *2* were logically combined, and holes were filled by first dilating the mask and then performing erosion from the outer boundary to obtain *mask 3*. This resultant mask was used for the BFR step (Supp. Fig. 1).

We applied each of the above-mentioned BFR algorithms to the unwrapped phase.

<u>SHARP and V-SHARP</u>: We chose resolution-independent high-pass regularization for the SHARP[29] and V-SHARP[30] techniques as recommended by Özbay et al.[34], along with the optimal radius ($R$=18vox; 1:18vox for VSHARP) and regularization parameter ($r$=0.0074) determined in that work.

<u>PDF and LBV</u>: PDF[33] and LBV[28] were originally developed using transceiver-phase free field maps from multi-echo data (personal communication). However, our phase data inherently contained transceiver phase contributions. We found that the transceive phase interfered with the convergence of PDF method when the default algorithmic parameters were used (30 maximum iterations; 0.1 tolerance). In consultation with the author (Pascal Spincemaille, Weill Cornell Medicine) of the PDF method, we optimized parameters (1500 maximum iterations; 0.1 tolerance) for our dataset and added 4th-order 3D polynomial fitting after PDF to suppress non-harmonic transceive phase contributions. The polynomial fitting was also applied after LBV with the same parameters as applied after PDF.

<u>RESHARP</u>: Since the regularization parameter specified in the original publication ($\lambda$=5·10$^{-3}$) resulted in obvious over-regularization we consulted with two of the original authors (Alan Wilman, University of Alberta; Hongfu Sun, The University of Queensland), who advised using

$\lambda=1\cdot10^{-4}$. Additionally, as per the authors' request, we applied polynomial fitting with the same parameters as used for LBV and PDF to the RESHARP output.

*Dipole inversion*

Each BFR algorithm created an eroded mask specific to the algorithm and subject scan (*mask 4*). We logically combined *masks 2* and *4* to create a mask for the dipole inversion step, *mask 5* (Supp. Fig. 1)

We applied each of the above-mentioned inversion algorithms to each of the BFR field maps. Their parameters can be found in the shared code (referenced above). For all algorithms, BFR maps were rescaled to the physical units expected by each algorithm before the algorithm was applied.

We received assistance from the authors of WH-FANSI, L1-QSM, HD-QSM (Carlos Milovic, Pontificia Universidad Católica de Valparaíso), and AMP-PE (Shuai Huang, Emory University) algorithms due to unsatisfactory results with the default parameters. For AMP-PE, the author recommended using high-order ('db1') instead of the default low-order ('db2') wavelet bases. Regarding WH-FANSI, we performed L-curve optimization under the guidance of the author. For L1-QSM and HD-QSM, although no exact parameters were suggested, the information provided by the authors allowed us to compute satisfactory susceptibility maps from these algorithms.

All computed susceptibility maps were multiplied with *mask 6* (Supp. Fig. 1) which was obtained by filling holes (as above) in *mask 5* before reporting the findings.

<u>Deep Learning Algorithms:</u> The included deep learning algorithms had specific requirements for the imaging slab orientation (all: strictly axial), voxel aspect ratio (all: isotropic), and voxel edge length (DeepQSM, QSMnet+: 1mm). For DeepQSM*, we used the native lateral resolution of 0.5mm. We applied the following pre- and post-processing steps to the background-field corrected phase images for these algorithms:

First, the background corrected field maps were spatially resampled (mri_convert, cubic) to the expected isotropic resolutions. Subsequently, the isotropic maps were rotated to strictly axial orientation using an in-house developed and thoroughly validated rotation tool based on FSL-FMRIB's Linear Image Registration Tool (FLIRT)[54] with trilinear interpolation using header information about the orientation of the imaging slab. We then applied the deep learning models and rotated the resulting susceptibility maps to their original orientation (native subject space). This processing was followed by resampling back to the original image resolution.

*Computation time*

We measured the average computation time of each inversion algorithm by applying it successively to all 396 BFR field maps (66 exams × 6 BFRs × 1 inversion). No GPUs were allocated to ensure comparability across algorithms.

**Analysis**

We calculated mean susceptibility values in the four DGM regions using an in-house developed bi-parametric atlas-based segmentation technique[55], which relied on both susceptibility and T1-weighted imaging.

We preprocessed the QSM and T1w images for multi-contrast template reconstruction as described earlier[55] (fully automated using pi4s). For this, we used susceptibility maps calculated with SDI[47] because it is one of the simplest QSM algorithms yielding susceptibility maps with relatively high visual quality, and it does not involve spatial regularization that could bias anatomical contrast in the final template. Third, a trained image analyst (A.Ad.) created atlas labels by manually outlining the four bilateral subcortical regions as well as the CSF (for referencing) in the template space using MRIcron software (v1.0.20190902) using both QSM and T1w contrast. We used SynthSeg[56] on the T1w template to obtain WM labels for referencing. Fourth, we propagated all atlas labels to the native subject spaces utilizing bi-modal warp field computations from Advanced Normalization Tools (ANTs).[57] Template-level labels and all propagated labels were carefully inspected by trained analysts (F.Sa., 5 years of neuroimaging experience; F.Sc, 15 years; N.Be., 20 years) and corrected where needed. WM labels were adjusted to accommodate the maximum erosion performed across all BFR algorithms (always the SHARP mask), which ensured consistency in voxel coverage of WM average values across all pipelines. Finally, we applied the subject-space labels to all reconstructed susceptibility maps and calculated regional mean susceptibility values in each hemisphere using FSLstats[58] (fully scripted). Subsequently, we averaged the regional mean values for both hemispheres to obtain an average regional mean susceptibility value, and referenced to each of the three reference regions.

**Statistical Analysis**

Statistical analyses were conducted using the Statistical Product and Service Solutions (SPSS; version 28; IBM, Armonk, NY) and Microsoft Excel (v16.8, Microsoft Corporation, Redmond, WA). Bi-lateral ROI measures were averaged between both hemispheres (fully scripted) and tested for normality using Q-Q plots and the Shapiro-Wilk test. Subsequently, paired t-tests were performed to test for inter/intra-algorithm and overall pipeline differences. P-values were corrected for multiple comparisons using the false discovery rate method (denoted as q-values) (Benjamini and Hochberg, 1995)[59] and results were considered statistically significant when $q<0.05$; uncorrected $p<0.05$ was considered a trend.

**Scientific Rigor - Analysis**

We conducted all analyses in a blinded manner, with unblinded metrics only being inspected collectively once metrics were available for all algorithms. Specifically, we refrained from comparing the reproducibility error and sensitivity values between algorithms until all metrics were available.

Subsequently, extensive quality control measures were implemented. This involved the visualization of randomly selected subject scans from various BFRs and inversion algorithms. The quality control process extended beyond the assessment of susceptibility maps alone; meticulous checks were performed on regional values to ensure accuracy.

**Correlation of Putative Over-time Iron Changes With Susceptibility Changes**

For each subject, we calculated putative regional iron concentrations from the regional age-dependency equations provided by H&S[22] using age at baseline and follow-up time points. Following this, we calculated the putative baseline-to-follow-up change in group-average iron concentrations for the enrolled cohort. Since the work by H&S did not provide an analytical trajectory for iron in the thalamus, we extracted the data from Fig. 8 in the publication[22] using WebPlotDigitizer (v3.9; Ankit Rohatgi, Austin, Texas, USA; http://arohatgi.info/WebPlotDigitizer/) and fitted a linear function to the data for ages 35 years and older.

We determined how well over-time susceptibility changes correlated with putative H&S iron changes by converting each subject's regional susceptibility values at follow-up and baseline to iron concentrations using a published conversion factor (13.2 ppb/mg iron/100g-wet-weight for ferritin at 36.5°C; regression equation specific to the gray matter).[3] We calculated the Pearson correlation between putative H&S iron changes and the susceptibility-derived iron concentrations. If Pearson correlation reached significance (p<0.05), we performed a linear least squares regression and recorded the slope value, which was expected to equal 1 if the iron concentrations derived from observed susceptibility changes quantitatively matched H&S putative iron changes. Finally, $R^2$ values were classified using Cohen's (1992) guidelines: ≤0.12 indicates low effect size, 0.13 to 0.25 indicates medium, and ≥0.26 indicates high effect size.[86] We repeated the analysis for iron values at baseline timepoint.

**Reproducibility**

We used the scan-rescan data to quantify each pipeline's reproducibility. Due to the short interval between the reproducibility scans, it can be assumed that $\sigma_{true,R}^2 + \sigma_{reference}^2 = 0$ in Eq. 2 and, hence, $\sigma_{overall,R} = \sigma_{measure,R}$, i.e. the *observed* scan-rescan variation is an estimate of the measurement-related variation.

We assessed a pipeline $p$'s reproducibility in ROI $r$ through a reproducibility metric, calculated from the subject-average scan-rescan standard deviation, $\hat{\sigma}_{overall,rp}$. Since dipole inversion algorithms have a propensity to underestimate magnetic susceptibility[29,72-75] (cf. Fig. 1), which would affect $\hat{\sigma}_{overall,rp}$ and, hence, would hamper the comparison of the metric across pipelines, we normalized $\hat{\sigma}_{overall,rp}$ with the average of the susceptibility across all four investigated DGM regions and all five subjects:

$$\rho_{rp} = \frac{\hat{\sigma}_{overall,rp}}{\frac{1}{5}\sum_{n=1}^{5}(\frac{1}{4}\sum_{r=1}^{4}\chi_{rpn})} \cdot 100, \tag{4}$$

where $\chi_{rpn}$ is the average susceptibility in region $r$ calculated with pipeline $p$ in subject $n$. The resulting reproducibility error metric, $\rho_{rp}$, denotes the amount of scan-rescan variation as a fraction of the average DGM intensity (in percent).

For visualization purposes, we also calculated the voxel-wise reproducibility error using susceptibility maps co-registered to the QSM-T1w template as:

$$\rho'_{pj} = \frac{\hat{\sigma}_{overall,rj}}{\frac{1}{5}\sum_{n=1}^{5}(\frac{1}{4}\sum_{r=1}^{4}\chi_{rpn})} \cdot 100, \quad (5)$$

where $j$ denotes the $j^{th}$ voxel.

**Sensitivity**

The practical relevance of scan-rescan reproducibility is limited without also considering the ability to detect change. A pipeline that would output the exact same susceptibility map independent of the input phase images would have perfect reproducibility (zero error or variation) but would not detect any changes in susceptibility over time or between individuals or groups (zero sensitivity). We used the absolute effect size, $|d_r|$, according to Eq. 1, to assess the ability of a pipeline to detect regional susceptibility differences, referred to as the pipeline's "sensitivity" in the following. When pipelines detected over-time changes inconsistent with H&S, signed effect sizes were considered. This distinction aimed to differentiate pipelines detecting over-time susceptibility changes in line and not in line with H&S.

For each pipeline $p$, we used the group average over-time susceptibility change of susceptibility in region $r$ as the nominator, $\Delta\chi_{rp}$, in Eq. 1, and the subject-average scan-rescan standard deviation, $\sigma_{measure,rp} = \hat{\sigma}_{overall,rp}$, as the denominator:

$$\hat{d}_{rp} = \left|\frac{\Delta\chi_{rp}}{\hat{\sigma}_{measure,rp}}\right|. \quad (6)$$

**Global Performance Metrics**

To capture and compare the average performance of pipelines across all DGM regions, we combined each pipeline's regional performance metrics, $\rho_{rp}$ and $\hat{d}_{rp}$, into a composite, global performance metric, $P$. We computed $P$ by averaging the regional findings, as

$$P_\rho = \frac{\sum_r^4 \rho_{rp}}{4} \quad \text{and} \quad P_d = \frac{\sum_r d_{rp}}{4}, \quad (7)$$

respectively.

**Percentile-based Classification**

To evaluate the distribution across pipelines, we computed the 33$^{rd}$ and 66$^{th}$ percentiles for the reproducibility and sensitivity metrics using the *prctile* function in MATLAB. The sensitivity of pipelines in the 33$^{rd}$ percentile was considered as *low*, in the 66$^{th}$ percentile as *medium*, and above the 66$^{th}$ percentile as *high*.

**Visualization**

We visualized all pipeline and region-specific metrics as heat maps using the seaborn library.[108] To facilitate visual inspection, we decided to mask (gray boxes in regional over-time and sensitivity; translucent in global sensitivity) any observations that deviated from the aging-related H&S-based over-time changes.

**Visual Evaluation of Top-Performing Pipelines**

We conducted a rater-based evaluation to qualitatively assess the quality of susceptibility maps of 10 subjects randomly selected from both the cohorts (longitudinal and reproducibility) in the 95$^{th}$ percentile of the sensitivity metric with three independent raters with 3 to 9 years of experience in brain QSM research (J.Re, T.Jo, and J.Ch). For increased scientific rigor, we included one rater (J.Ch) who had previously worked in the Cornell MRI Research lab (Weill Cornell Medicine) and had no prior experience with our in-house developed algorithms and the appearance of the resulting susceptibility maps. The other two raters obtained the majority of their experience with QSM in our lab (J.Re, T.Jo). Raters were blinded to the pipelines used. The evaluation was conducted using ITK-SNAP[60], with all three 95$^{th}$ percentile maps presented side-by-side with the same preset contrast setting. For each subject, raters were instructed to rank the presented susceptibility maps, considering the following criteria: Presence of artifacts (e.g., streaking); signal homogeneity; visibility and differentiation of anatomical (fine) structures; spatially varying intensity inhomogeneities; overall natural appearance of the image. Regional intensity differences were not considered for ranking. One of the raters (J.Re) was selected for an intra-rater evaluation after two weeks of the initial ranking. The susceptibility maps were shuffled and the same procedure described previously was repeated for ranking.

Fleiss' kappa[61] was calculated to assess the inter-rater agreement and assessed using the classification scale of kappa coefficient ($\kappa$) designed under the guidelines from Altman.[62] Additionally, the intra-class correlation coefficient was assessed in accordance with established guidelines from the literature.[106]

# Results

**Participants**

Two subjects had to be excluded from the study due to data integrity issues, resulting in N=23 subjects for the longitudinal analysis. The average age of included subjects was 57±9 (39-73) years at the time of the baseline scan with a female:male ratio of 19:4. The median time between baseline and follow-up scans was 10.0 years [11.5-13 IQR].

**Computation and Quality Control**

Times of dipole inversion computations varied between 22 seconds (TKD) and 46 hours (WH-FANSI) per scan (Supp. Fig. 3).

Reconstruction failed with DeepQSM (all BFRs) for one follow-up scan of one subject. Additionally, one scan at baseline, computed using LBV+FANSI, was excluded due to failed reconstruction. Consequently, their respective baseline and follow-up scans were excluded from the subject-wise H&S-based correlations (Fig. 5) performed to assess over-time changes. Exclusion did not alter the median time between the baseline and follow-up scans (10 years [11.5-13 IQR]) used to investigate the longitudinal susceptibility changes in this study. Figure 1 illustrates susceptibility maps resulting from all inversion algorithms. The visual appearance of the susceptibility maps was highly similar, and no major artifacts were visible on any of the maps. Maps differed primarily in intensity (related to systematic underestimation) and edge definition (smoothing), two well-characterized limitations of many algorithms.[17,18]

**Reproducibility**

The choice of the BFR algorithm had a systematic effect on the reproducibility, with some BFR algorithms performing systematically better than others, independent of the dipole inversion algorithm. Figure 2 summarizes reproducibility error metrics for each pipeline and region.

<u>Pipelines with SHARP-based BFRs demonstrated the best reproducibility.</u>

Irrespective of the inversion algorithm and reference region, SHARP-based BFRs stood out for consistently high reproducibility (low reproducibility error) in all anatomical regions except for the GP. Using either WB or WM reference regions, 18 out of 21 (86%) inversion algorithms achieved their highest reproducibility with RESHARP (exception: WH-FANSI, DeepQSM and L1-QSM).

We compared BFRs statistically by calculating the median reproducibility error for each algorithm and conducting a paired t-test analysis. For WB and WM referencing, we compared SHARP to all other BFRs because it was highly reproducible independent of the inversion algorithm with either WB (median=5.9 [1.1 IQR]) or WM referencing (5.1 [0.8]). All BFRs, except for RESHARP (q=0.50), displayed significantly lower reproducibility (high reproducibility error) compared to SHARP (q≤0.026). We used VSHARP as the benchmark for comparison of CSF-referenced pipelines because this BFR exhibited the lowest reproducibility error independent of the inversion algorithm (median=8.6 [1.8]) with CSF referencing. RESHARP and SHARP demonstrated comparable results to VSHARP (q=0.60), while other BFRs exhibited significantly higher reproducibility errors (q=0.002).

Inversion with QSMnet+ demonstrated the best reproducibility

In all regions except for the GP, QSMnet+ consistently ranked among the best in reproducibility, regardless of the chosen BFR and reference region.

CSF referencing resulted in poor and WM referencing in best reproducibility

Significantly improved reproducibility was observed when using the WB (35% reduced reproducibility error, q=0.02; paired t-test) or WM (43% reduced reproducibility error, q=0.003) as a reference region, compared to CSF. The median normalized reproducibility error across all pipelines and regions was 5.90 [4.25] for WM reference, 6.65 [4.82] for WB reference, and 10.30 [5.32] for CSF reference. No significant difference was observed between WB and WM reproducibility (q=0.07; paired t-test).

Changing the reference region from WB to WM reduced reproducibility error between 8% (caudate) and 15% (thalamus). Changing from WB to CSF increased the reproducibility error between 36% (GP) and 100% (putamen). See Supplementary Figure 7 for a quantitative overview of the impact of the choice of the reference region on reproducibility using RESHARP as an example BFR (chosen because of its high reproducibility, Fig. 2).

Reproducibility was similar or worse compared to previous investigations

Employing similar pipelines as previous investigations at 3T[63-65], the present study demonstrated lower or comparable scan-rescan variability (Supp. Fig. 8). However, as anticipated, the variations were higher compared to those observed at 7T.[66]

GP presented with the worst reproducibility across all regions

Figure 3 illustrates the effect of the reference region on voxel-wise reproducibility error for selected inversion algorithms. We chose VSHARP as a representative BFR algorithm due to the substantial range of reproducibility error values observed with this algorithm. Referencing-related changes in reproducibility error and differences between inversion algorithms were widespread and not localized to specific brain regions. Voxels in the GP region consistently demonstrated the highest reproducibility error compared to other regions ($q<0.001$), in line with the ROI analysis (see above; Fig. 2).

Global reproducibility

Figure 4 summarizes the data in Fig. 2 using the global reproducibility error metric (Eq. 7). The global region-averaged reproducibility error metrics largely mirrored the regional analyses (Fig. 2). Global reproducibility error varied from 4.0 (SHARP or RESHARP; LSQR; WM) to 23.8 (LBV; DeepQSM*; WB).

Highest reproducibility with QSMnet+ (WB), LSQR (WM), and SDI (CSF) inversion

We chose QSMnet+ as a benchmark for comparison with other inversion algorithms because this algorithm demonstrated low reproducibility error with WB referencing across all BFRs (median=5.10 [2.05]). Independent of the BFR, 50% of the inversion algorithms demonstrated significantly higher reproducibility errors than QSMnet+ (q≤0.042; L1-QSM: median=11.61 [3.53], DeepQSM: 10.8 [2.93], HD-QSM: 9.40 [2.82], FANSI: 9.05 [3.50], WH-FANSI: 8.04 [1.13], MEDI: 6.93 [2.05], MEDI+0: 6.69 [1.71], MEDI+0*: 6.58 [2.14], iSWIM: 6.48 [2.10], MATV: 6.46 [2.13]). Differences in reproducibility did not reach significance for the remaining algorithms (q≥0.06) across all BFRs.

With WM-referencing, LSQR's reproducibility was observed to be the best, independent of the BFR (4.65 [0.97]). Significantly higher reproducibility error compared to LSQR (q≤0.04) was observed with L1-QSM (10.08 [2.91]), DeepQSM (9.75 [2.82]), FANSI (7.99 [3.36]), WH-FANSI (7.02 [0.81]), iLSQR (6.29 [1.57]), DirTIK (5.64 [0.24]), and TKD (5.61 [0.48]). The remaining algorithms did not reach significance (q≥0.06).

For CSF-referencing, SDI exhibited the lowest median reproducibility error (7.48 [1.50]) independent of the BFR. Differences reached significance for all algorithms (q≤0.05) except for IterTIK and QSMnet+ (q≥0.3).

**Over-time Changes**

For conciseness of the presentation, we focused the presentation of results to WB-referenced findings (Supp. Figs. 10 and 11) because of its superior performance compared to CSF in the previous analyses and generally similar findings compared to WM referencing (Supp. Figs. 12 and 13 for WM; Supp. Figs. 14 and 15 for CSF-referenced findings).

Inter-pipeline variability in over-time changes is not related to systematic underestimation

Detected over-time changes varied by up to two orders of magnitude between algorithms with changes ranging from -0.05 to -19 ppb in the thalamus, +3.1 to +34 ppb in putamen, +1.2 to +40 ppb in caudate, and +0.18 to +42ppb in GP. Normalization of the values to the total DGM, to compensate for systematic underestimation, reduced some outliers (such as for DeepQSM) but did not reduce the overall variation in detected changes: -1 to -304 in thalamus, +22 to +459 in caudate, +35 to +459 in putamen, and +2 to +481 in GP.

High correlation with and overestimation of putative over-time iron changes

Calculated group-average putative over-time changes in iron concentrations according to H&S were (in increasing order): -0.347 of fresh tissue weight (mg/100g) in thalamus ($c_{Fe}(age[y]) = -0.0358 \cdot age[y] + 6.9102$), +0.101 mg/100g in GP, +0.237 mg/100g in caudate, and +0.512 mg/100g in putamen.

Figure 5 depicts the coefficients of determination, $R^2$, (top block) for detected over-time susceptibility changes in the DGM (WB-referenced) and the putative iron changes according to H&S's region-dependent equations. Additionally, the plot shows the associated slopes (bottom block) for each pipeline, indicating whether the pipeline overestimates (>1) or underestimates (<1) the susceptibility changes compared to the presumed changes based on H&S. All but two pipelines demonstrated a significant linear correlation ($p<0.05$, $R^2$ threshold=0.02). Out of the remaining 124 pipelines, 71 pipelines (57%) overestimated putative iron concentrations by at least a factor of two. DirTIK with PDF (slope=1.04), SDI with VSHARP (1.12), RESHARP

(1.13), LBV (1.07) and PDF (0.87), and iLSQR with RESHARP (1.07), estimated over-time changes close to the putative iron changes (slope between 0.9 and 1.1).

Putative iron changes according to H&S explained between $R^2=1\%$ (PDF+DeepQSM and LBV+DeepQSM*) and 39% (RESHARP+MATV) of the variation in susceptibility values. 19 pipelines exhibited low correlations ($R^2 \leq 0.12$), 63 exhibited medium correlations ($\geq 0.13$ and $\leq 0.25$), while the rest (44) showed high correlations ($\geq 0.26$). Seven pipelines stood out within the 95th percentile ($\geq 0.35$): RESHARP with MATV, AMP-PE, LSQR, MEDI+0, iSWIM, and MEDI+0*, and SHARP with MEDI+0 (in descending order).

Iron values at baseline consistenly showed high correlation ($R^2 \geq 0.76$) for all but one pipeline ($N$=125/126; 99.2%). 62 of 126 pipelines (49%) estimated iron values close to the putative iron values (slope between 0.9 and 1.1), while the remaining pipelines either overestimated (maximum of +34% - DeepQSM with LBV) or underestimated (-68% - DeepQSM* with LBV) iron values. Results are visualized in Supplementary Figure 16.

<u>Pipelines inconsistent with H&S predictions</u>

To examine inconsistencies in voxel-wise over-time changes, particularly those deviating from predictions based on H&S (most of PDF- and iHARP-based pipelines), we conducted a post-hoc analysis across all pipelines, depicted in Fig. 6. Upon scrutinizing these maps in axial, sagittal and coronal views, we found that iHARP did not reveal any specific reason for the observed decline in GP susceptibility. Conversely, PDF displayed increasing thalamic susceptibility, possibly due to long-ranging hyper-intense artifacts co-locating with the DGM regions. Notably, QSMnet+ did not exhibit such artifacts with any of the BFR methods.

The regional over-time change detected with PDF, in the absence of artifacts within the DGM regions, is comparable to the findings from other BFRs (for e.g., PDF+SDI=153 ppb compared to RESHARP+SDI=101 ppb change in putamen, and PDF+QSMnet+=250 ppb compared to RESHARP+QSMnet+=246 ppb). However, the presence of artifacts (Fig. 6 - last row, left column) amplified the change (putamen: iterTik=332 ppb) that does not fall within the comparable range to other BFR over-time change using the same inversion algorithm (second highest change observed from SHARP=138 ppb).

No prominent artifacts were observed in maps from LBV and SHARP-based BFRs (select RESHARP-based inversion maps displayed in first row in Fig. 6).

<u>DeepQSM detected over-time change in the thalamus inconsistent with the H&S prediction</u>

In the thalamus, the DeepQSM inversion algorithm did not detect the presumed over-time changes with any of the BFR algorithms tested.

<u>Differences in detected over-time changes between BFRs and inversion algorithms</u>

Within the putamen region, all pipelines detected over-time susceptibility changes signed consistently with H&S, a region with the highest change detected with most pipelines. Focusing specifically on the putamen region, we replicated the inter-algorithmic paired t-test analysis previously conducted within the reproducibility error global metric for the normalized over-time change findings.

Within the putamen region, PDF detected the highest median change (272.2 [86.2 IQR]) across all BFRs. However, our voxel-wise over-time change analysis (Fig. 6) revealed substantial within-DGM artifacts associated with PDF. Consequently, we chose RESHARP (204.24 [62.2]), which had the second-highest median over-time change, for comparison with other BFRs. LBV and VSHARP exhibited significantly lower over-time changes (q≤0.01), while PDF showed higher over-time change (q=0.003) compared to RESHARP. Differences in over-time changes between SHARP and iHARP did not reach significance in comparison to RESHARP (q≥0.30).

Regarding inversion algorithms, DeepQSM (median across BFRs=265.9 [57.7] - highest in putamen) was compared to all other algorithms. No significant difference was observed for FANSI, WH-FANSI, DeepQSM*, L1-QSM and HD-QSM (q≥0.10), while all others detected significantly lower changes over time compared to DeepQSM (q≤0.05).

**Sensitivity**

Figure 7 represents sensitivity findings using WB referencing. WM and CSF-referenced findings can be found in the Supplemental Materials (Supp. Fig. 21 and 22, respectively).

<u>Sensitivity varied between anatomical regions</u>

Sensitivity significantly varied between DGM regions (q≤0.02) with the highest median sensitivity within the putamen region (5.04 [3.49 IQR]), as expected based on the high putative over-time iron changes (see above). The putamen was followed by caudate (3.13 [1.38]), thalamus (2.25 [1.25]), and GP (1.20 [1.36]).

<u>Pipelines with RESHARP and PDF demonstrated the highest sensitivity</u>

Most pipelines yielded medium to high sensitivity values in all regions, but sensitivities varied substantially between pipelines within each region. None of the pipelines were consistently better than other pipelines across all regions. RESHARP demonstrated high sensitivities with 62% of the inversion algorithms (13 of 21) in the thalamus. In the remaining regions, PDF yielded higher sensitivities with most inversion algorithms (10, 11 and 9 out of 21 inversions in GP, caudate and putamen, respectively) compared to RESHARP.

Within putamen, pipelines with RESHARP and PDF yielded similar median sensitivity (7), which was on average 36% higher than pipelines using other BFRs. Within the GP region, PDF yielded the highest median sensitivity of 3.7 (on average 70% higher), while, within the caudate, all BFRs were similar with PDF ranking the highest (3.9). Within the thalamus, RESHARP ranked highest with a median sensitivity of 3.85. Figure 8 summarizes these findings.

High sensitivity across all regions with RESHARP

Figure 9 illustrates the global pipeline sensitivity metrics. Results largely mirrored the reproducibility error findings (Fig. 4), particularly regarding the impact of the reference regions. WB and the WM-referencing exhibited similar effects on the overall DGM sensitivity ($p=0.50$), whereas CSF-referencing markedly reduced sensitivity ($p<0.001$) compared to WB and WM-referencing.

The combination of RESHARP with AMP-PE yielded the overall highest sensitivities with either WB [$P_d$=5.49] or WM referencing [5.62], followed by RESHARP with HEIDI, and RESHARP with LSQR (all $P_d \geq 5.00$ - excluding PDF-based pipelines due to artifacts shown in Fig. 6).

With CSF referencing, the highest sensitivities were observed when RESHARP was used with HEIDI, SHARP with HEIDI, SHARP with one of the variations of MEDI+0, and iHARP with QSMnet+ (all $P_d \geq 3.00$).

Overall, SHARP-based BFRs (SHARP, VSHARP and RESHARP) outperformed other BFR algorithms with respect to sensitivity and the number of pipelines (19 each for SHARP and RESHARP, 18 for VSHARP) that detected over-time changes consistent with H&S (excluding PDF because of artifacts). RESHARP-based pipelines exhibited the highest sensitivities across all BFRs (median across inversion algorithms: 3.74 [1.48] for WB, 3.46 [1.62] for WM, 1.37 [2.29] for CSF), except for pipelines with deep learning-based inversion algorithms, which either yielded H&S-inconsistent results (DeepQSM and QSMnet+) or marginally better results with SHARP or VSHARP (DeepQSM*).

iLSQR was the only inversion algorithm detecting longitudinal changes consistent with H&S regardless of BFR and reference region. Additionally, iSWIM, STAR, LSQR and L1-QSM also demonstrated similar consistency in terms of H&S-consistent temporal changes in all regions with WB and WM-referencing.

**Inter- and Intra-Rater Evaluation**

After evaluating all pipelines based on the WB-referenced global sensitivity metric (top panel - Fig. 9), five pipelines emerged with sensitivities in the 95$^{th}$ percentile ($P_d \geq 4.59$), PDF with L1-QSM (5.45), RESHARP with AMPPE (5.49), RESHARP with HEIDI (5.38), RESHARP with LSQR (4.97), and PDF with HEIDI (4.60). However, due to presence of artifacts within the PDF-based reconstructions (see Fig. 6), we excluded pipelines with PDF from the following evaluation.

All raters ranked the combination of RESHARP and HEIDI the highest (ranked 1$^{st}$ in 29 of 30 ratings; 97%, see Supp. Fig. 24). Inter-rater agreement value ($\kappa$) was 0.28 (p<0.005; fair agreement; primarily driven by mixed ranking for LSQR and AMP-PE). Intra-rater agreement was 0.90 (excellent).

<u>Raters preferred homogeneity and sharpness of HEIDI-based susceptibility maps</u>

We conducted a post-hoc evaluation with all raters two weeks after the second evaluation to understand why raters preferred susceptibility maps calculated with HEIDI over those from AMPPE and LSQR. Raters were blinded to the algorithm names during this evaluation. All raters responded that HEIDI's susceptibility maps were the sharpest and most homogeneous, while LSQR exhibited obvious streaking artifacts, inhomogeneity, and more noise overall. The AMP-PE maps were criticized for being blocky (pixelated when zoomed in), having lower boundary definition, and suffering from blurry streaking artifacts. Raters' individual comments can be viewed in the Supplemental Materials (Supp. Table 1).

# Discussion

This study systematically investigated QSM pipelines concerning scan-rescan reproducibility error and sensitivity toward over-time changes in brain susceptibility.

**Study Design**

The fundamental premise of this study was that susceptibility changes due to aging are dominated by alterations in tissue iron, as described half a century ago by H&S.[22] Over-time changes in the present study can be considered as a surrogate for an algorithm's general ability to detect changes in susceptibility between two sample distributions, such as between a patient and a control group in a clinical study setting.

The considerable variation in observed over-time susceptibility changes across pipelines suggests that it may be challenging to quantitatively compare findings of longitudinal studies that used different pipelines. Despite this variability, most pipelines demonstrated medium to high quantitative correlations in over-time susceptibility changes with putative iron changes (Fig. 5 – top panel). However, while correlations were high, the estimated QSM-dependent changes were at least twice as large as presumed (Fig. 5 – bottom panel). Hence, qualitative study outcomes, such as the statistical significance of group differences, are likely reproducible across pipelines, but caution should be applied when comparing outcome measures across studies that used different pipelines. Future research needs to investigate the effect of pipeline choice on the outcomes of cross-sectional and longitudinal studies and whether findings previously reported in the literature can be confirmed (reproduced) with different pipelines, group comparison and over-time study designs. The computational framework developed in the present study could be applied directly to other imaging datasets and may be useful to shed further insights on the pipeline dependence of QSM.

The reason for the substantial overestimation of aging-related iron changes relative to H&S remains unclear but may be attributed to the different environmental exposure and cultural and lifestyle circumstances of the cohort studied by H&S compared to our cohort, more than half a century later and in North America vs. Scandinavia. It is likely that our cohort experienced iron changes at a different rate than that studied by H&S. Lastly, while our inclusion criteria intentionally restricted the age range to a period of putatively stable myelin concentration, there is a remaining possibility that alterations in myelin, e.g. due to aging-related demyelination[84,85], affected the comparisons with H&S-derived iron concentrations and may have been responsible for H&S-inconsistent outcomes observed with some pipelines. However, we consider the contribution of demyelination to be low because it would have confounded the thalamic over-time decline, which was largely consistent with H&S.

The high consistency of *cross-sectional* putative iron values with QSM-based iron concentrations from most pipelines at baseline (slopes between 0.9 and 1.1 in Supp. Fig. 16, bottom) has to be interpreted with caution because the susceptibility-to-iron conversion factor in Langkammer et al.[3] was itself determined using QSM (HEIDI and SHARP). However, a slope close to one and very high correlation suggest that iron values reported by H&S are still representative for those found in today's population and, second, provides further support for the study's premise that DGM susceptibility is dominated by susceptibility effects from tissue iron.

**Region-Dependent Reproducibility**

Our finding of region-dependent variation in reproducibility error, with the highest error in the GP, was in line with previous reproducibility studies (Supp. Fig. 8).[63-66] While it was beyond the scope of our study to identify the methodological causes of scan-rescan variability, the absence of apparent contributors to variation in our voxel-wise analysis patterns (Fig. 3) suggest that random effects such as noise amplification and motion artifacts are the major contributors to scan-rescan variation.

Furthermore, consistent with a previous study[66], we contend that opting for a single-echo sequence (as in the present study) may provide lower reproducibility error of QSM, particularly within the DGM, compared to multi-echo sequence(s) used in previous studies[63,65], and recommended by the recent QSM Consensus.[21] Further research is needed to corroborate this observation.

**High Performance of SHARP-based BFRs**

BFR algorithms have not been evaluated with the same scrutiny as dipole inversion algorithms in the past[17,18,67,69,88] likely because it was assumed that their effect on susceptibility map reconstruction quality is small compared to that of the dipole inversion. Our observations confirm this notion. While we identified spatially slowly varying remnant fields (Fig. 6) as a confounder of scan-rescan reproducibility error, these remnant fields, induced or left uncorrected by the BFR algorithms, could only be visualized through difference images and were indiscernible on both the background-corrected field maps and the final susceptibility maps.

After correcting for systematic underestimation, regional reproducibility error was overall relatively similar across pipelines (Fig. 2). The slightly increased robustness of SHARP-based BFRs, also recommended in the 2023 QSM Consensus[21], may be related to the spatial averaging intrinsic to the spherical mean value computation within these methods, as well as mild low-pass filtering capabilities. The spherical mean value computation may be less sensitive to field errors and noise close to the brain's surface compared to other methods.[67] The lowpass filtering may suppress residual background fields or transceiver-phase contributions that interfere with the dipole inversion. Residual transceiver-phase contributions in our data may have amplified the importance of low-pass filtering in these algorithms and the difference between BFRs may be less pronounced when the transceive phase is removed analytically using multi-echo data.[21,31]

**Remarkable Reproducibility of QSMnet+**

QSMnet+ consistently achieved the highest reproducibility independent of the BFR (Fig. 4). This finding was in line with the original publication's assertion of high reproducibility[68], as they anticipated their network to deliver such performance based on consistent WM contrast observed in multiple head orientation QSM outputs from QSMnet. This analogy suggests that the reliability observed in WM contrast extends to the DGM. Using VSHARP and WB-referencing, as in the original publication, the reproducibility error found in the present work (caudate, putamen, and GP=0.040, 0.036, and 0.124 ppm, respectively) fell within a close range to that reported in the original publication of the precursor method, QSMNet[68] (0.034, 0.054, and 0.130 ppm, respectively; QSMnet+ publication[46] did not investigate reproducibility). While investigation of the reasons for the high reproducibility and BFR independence of QSMnet+ was beyond the scope of this study, the responsible features of the algorithm may hold the key to lower reproducibility error and higher sensitivity of QSM. For example, the low reproducibility error of WM measurements may have positively affected referencing-related variation.

**Complex Interplay Between BFR and Dipole Inversion**

The sensitivity (Fig. 7) toward over-time changes (Supp. Fig. 10) demonstrated a more profound interplay between BFR and dipole inversion algorithms than the reproducibility error metric. While mechanistic investigations were beyond the scope of the study, changes in brain anatomy and overall susceptibility distribution from baseline to follow-up are likely contributors to this interplay. While brain anatomy and susceptibility were identical between scans in the scan-rescan experiments, they were naturally affected by a decade of aging in our longitudinal experiments. BFR artifacts that are in some way linked to brain anatomy, e.g. through the BFR-intrinsic masks, through features of the background field itself, or through spatial-frequency filtering[67], can be expected to affect over-time changes whereas their effect on scan-rescan experiments might be negligible. This aspect of QSM reproducibility has wide-ranging implications of the use of QSM in clinical studies but is currently only poorly understood. The substantial variation in sensitivity between pipelines suggests that future QSM benchmarking initiatives should focus on full pipelines rather than single steps within the pipeline. We speculate that differences in dipole inversion algorithms' ability to handle BFR artifacts are likely related to the employed regularization approaches.

**Effect of Referencing on Reproducibility Error**

Our study confirmed the recent consensus statement[21] that larger reference regions, particularly WB referencing, yield more stable and reproducible susceptibility values than smaller regions. While CSF susceptibility is unaffected by most diseases, CSF referencing suffers from numerous challenges that were discussed previously[21], including a potential age dependency.[83] In particular, CSF-referencing resulted in the lowest reproducibility and sensitivity in the present study, although this finding conflicts with another study.[20]

While MEDI+0 addresses certain limitations of CSF-referencing by enforcing a homogeneity constraint for the CSF susceptibility, the CSF-constraint showed limited efficacy in improving reproducibility compared to the same method without the CSF-constraint or the conventional MEDI (2011) method.[44] This observation was in line with results presented in the original MEDI+0 publication (Fig. 5 in Liu, Z. et al., 2017).[45]

**Limitations**

Due to the extensive number of pipelines investigated, optimization of each algorithm's parameters was infeasible. We relied on default parameter settings, parameter settings recommended for our data, or consulted with the original developers if algorithm performance was considerably below that observed with other algorithms or presented in the original publication (see Methods). Consequently, it is possible that pipeline performance can be improved in some cases by parameter optimization. Since it is well-known that algorithmic parameter choices in the dipole inversion step substantially affect the appearance of computed susceptibility maps, and this has been confirmed by recent benchmarking studies[17,18], we performed a preliminary post-hoc analysis of the effect of dipole inversion regularization parameters. We computed WB-referenced reproducibility and sensitivity metrics for MEDI[44] with all BFRs across an extreme range of regularization parameters (Supp. Fig. 25). We chose MEDI because of the wide-spread use of the algorithm and its well-known sensitivity of the susceptibility map appearance on the regularization parameter. As expected, the appearance of the computed susceptibility maps differed greatly depending on the parameter values (supp. Fig. 26). However, despite these massive changes, the reproducibility and sensitivity metrics were largely unaffected. This counter-intuitive finding supports the generalizability of the present study and may be explained by the ROI-based calculation of the sensitivity and reproducibility metrics. Most regularization strategies penalize high-frequency image noise and streaking artifacts and prioritize edge delineation. Both effects, while effectively manipulating the visual appearance of the susceptibility maps and the conspicuity of subtle and small-scale susceptibility features, have a limited effect on the mean ROI values. This preliminary investigation suggests that lack of optimization of regularization parameters may not have substantially biased the present study or any clinical QSM study in the literature. Robustness of our findings with respect to regularization differences and the substantial differences in the appearance of the algorithms in the 95th sensitivity percentile challenge the practical relevance of past benchmarking initiatives for clinical research[17,18], which prioritized quantitative reconstruction accuracy with respect to a gold standard susceptibility map. Our data suggests that quantitive accuracy and visual quality do not imply high sensitivity toward susceptibility differences, and vice versa.

In this study, we did not investigate the effect of unwrapping algorithms on the final solution. We used best-path unwrapping due to its wide-spread use, absence of the structural modification of phase images[107], and because it is an exact phase unwrapping method recommended by the 2023 QSM consensus.[21] Including different unwrapping algorithms would have rendered the computational cost and complexity of the presentation of the results in this study infeasible. For the same reason, we also did not investigate the effect of the different brain mask creation algorithms. While our quality control did not reveal any issues within the resulting BFR maps, the mask creation may still have affected the BFR performance.

Another limitation of the study is the reliance on data acquired at a single site with a single pulse sequence. While several multicenter studies have shown good reproducibility of QSM across sites and acquisition pulse sequences[64,66,70,71], our findings may not fully generalize to all scanner and pulse sequence configurations. Specifically, our study used a single-echo sequence and transceiver phase contamination may have affected our results, although we did not find evidence supporting this hypothesis. While the 2023 QSM Consensus[21] recommends multi-echo acquisition, a recent study found that single-echo acquisition can be beneficial for QSM reproducibility.[66] Our sequence also used anisotropic voxel size which may have differential impact on algorithms. Due to the long follow-up time in the present study, it is challenging to confirm the generalizability of our results at a different site with different data acquisition.

# Conclusion

Our results highlight the importance of considering the performance of the entire QSM pipeline rather than that of individual components in isolation. While most of the 378 QSM pipelines included in this study reliably detected DGM iron changes over time, sensitivity varied substantially across pipelines, with BFR being a major contributor to variations in sensitivity.

# Figures

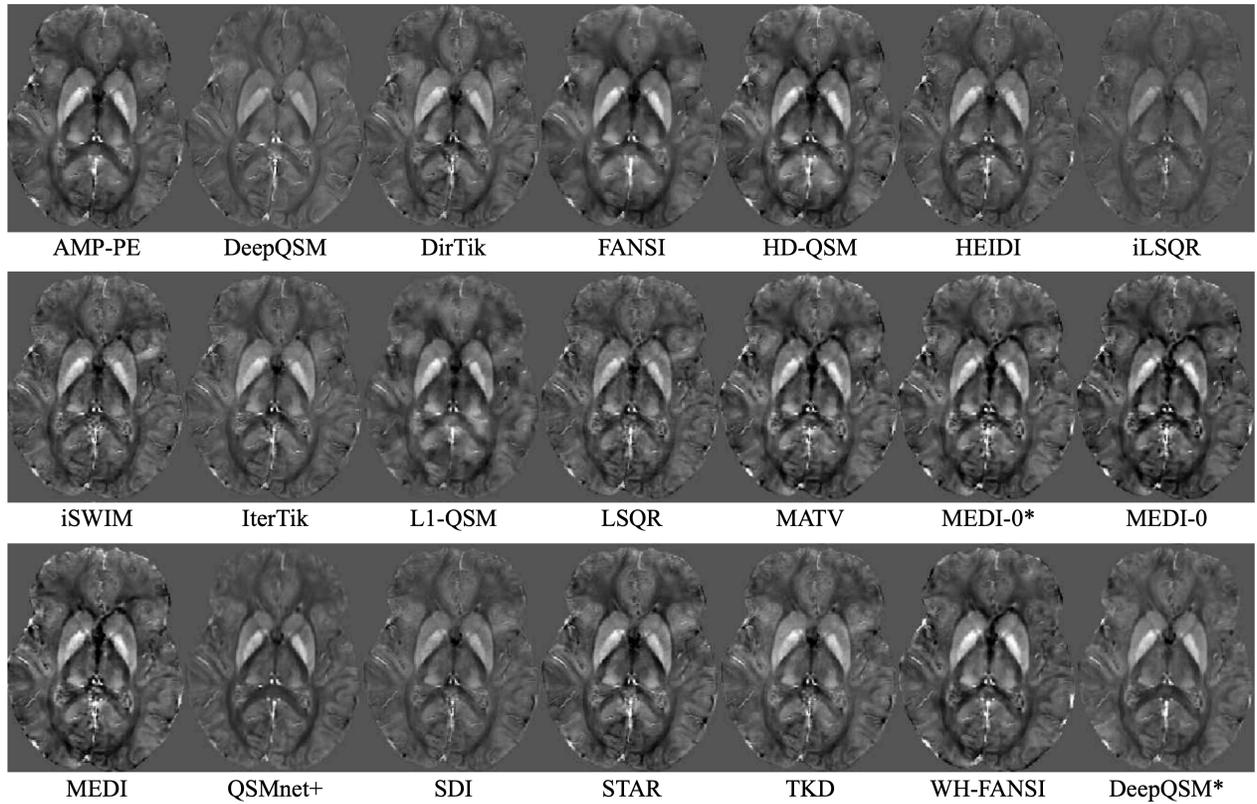

Fig. 1. Susceptibility maps in native space of a representative healthy subject (26 years old male) from different inversion algorithms using LBV as the BFR. Contrast range: -0.08 ppm (black) to 0.15 ppm (white).

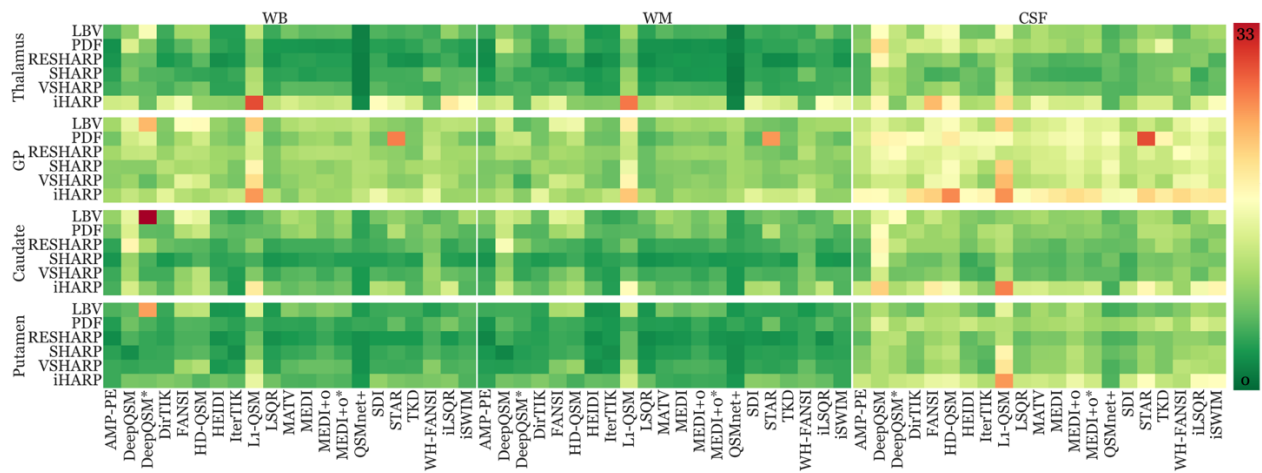

Fig. 2 Normalized reproducibility (scan-rescan variation) findings of each pipeline and region. Lower values (green) represent higher reproducibility (lower variation), and vice versa for red. Each group of three horizontal panels

combines the results of one specific DGM region that is denoted at the left-hand side of the panel. The three horizontal panels of each region summarize results using one specific reference region, which is listed at the top. Within each panel, each row corresponds to a specific BFR algorithm (listed on the left-hand side) while each column represents an inversion algorithm (listed at the bottom). See Supplemental Materials (Supp. Figs. 4-6) for an annotated version of the figure with individual pipeline metrics.

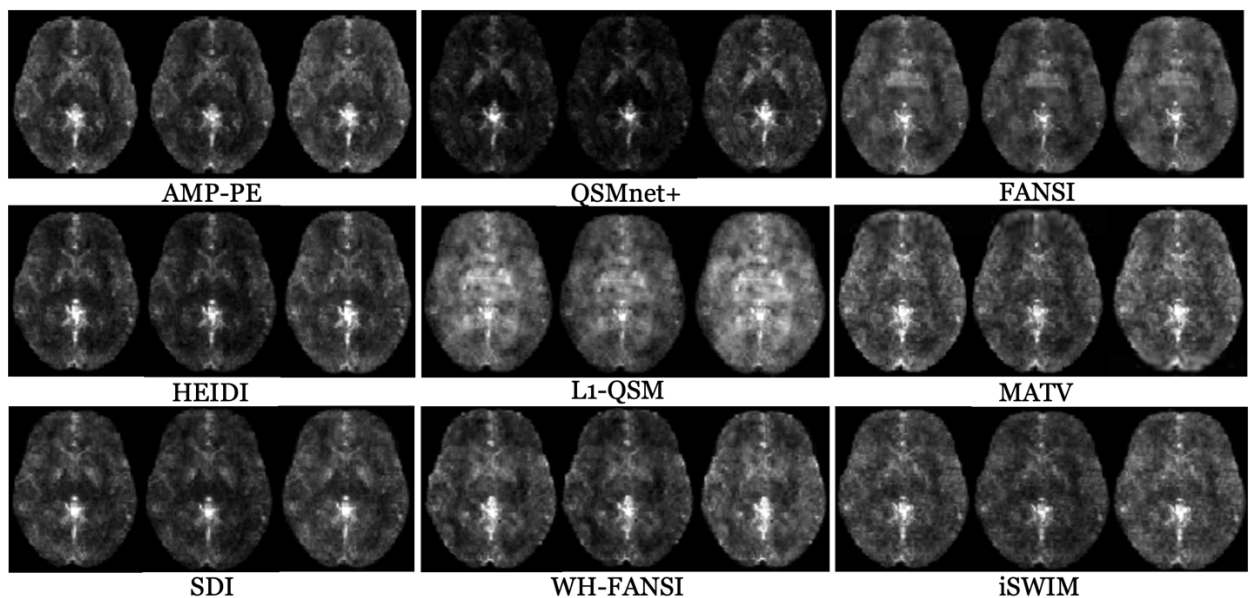

Fig. 3 Voxel-wise normalized reproducibility ($\rho_{rj}$) for V-SHARP with selected inversion algorithms covering a wide range of reproducibility values in the same slice of the template as shown in Supp. Fig. 2. From left to right, susceptibility maps were referenced to WB, WM, and the CSF, respectively. The contrast is set from 0 to 0.55 (no unit).

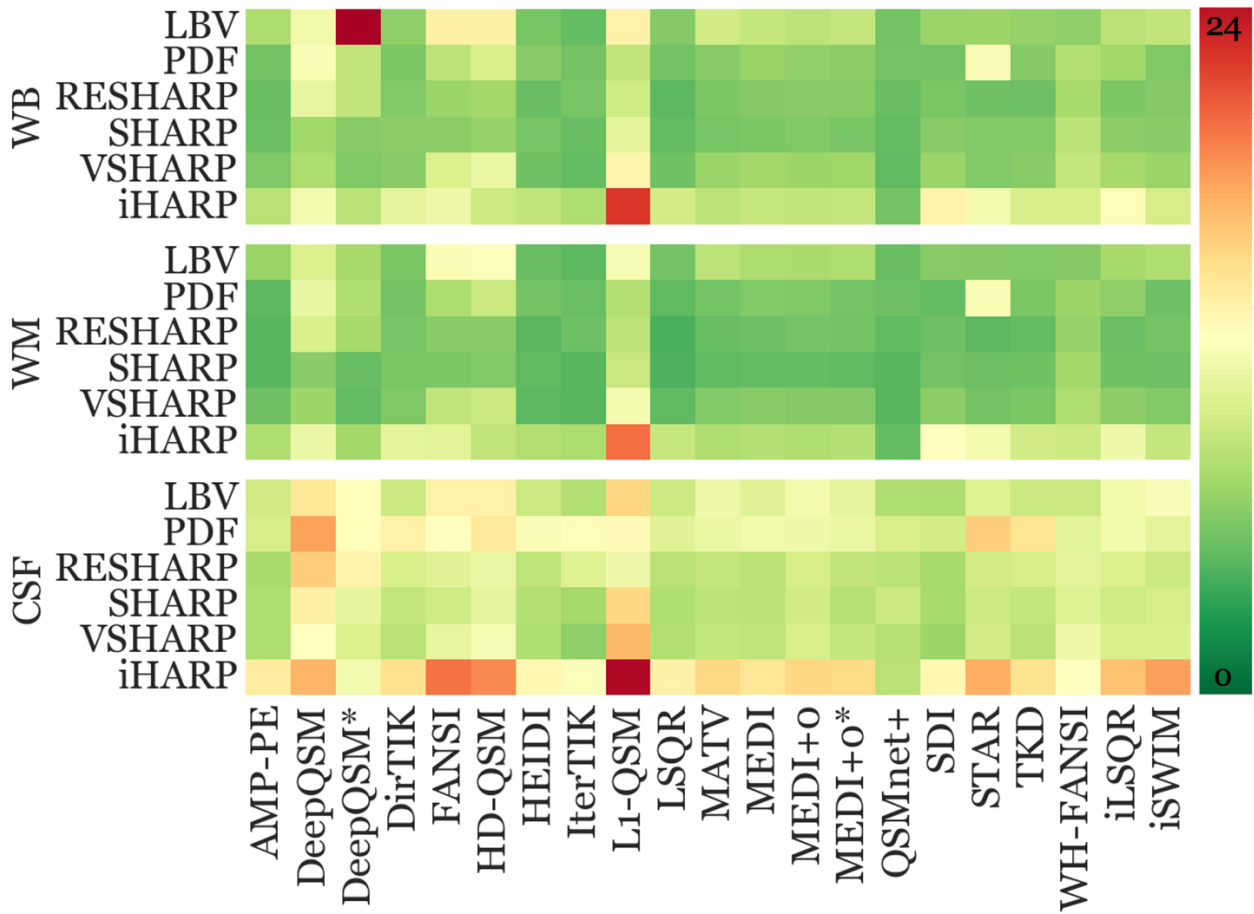

Fig. 4 Global performance reproducibility (scan-rescan variation) metric according to Eq. 7 (left). The color-coding and the arrangement of BFR and inversion algorithms mirrors that of Fig. 2. In this figure, each panel represents a specific reference region (listed on the left-hand side). See Supplemental Materials (Supp. Fig. 9) for an annotated version of this figure.

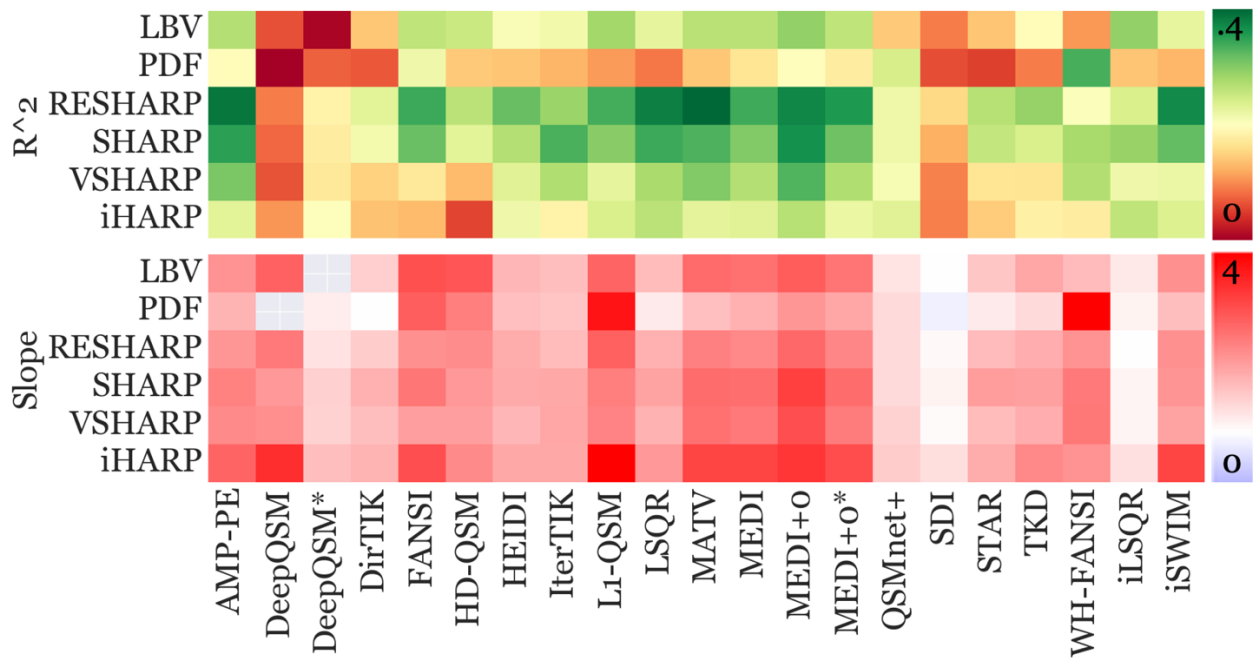

Figure 5. Correlation between observed DGM over-time susceptibility changes and putative iron changes. The coefficients of determination, $R^2$, are displayed at the top, while the slope (white box = 1) is depicted at the bottom. Pipelines within the slope plot that exhibited non-significant Pearson correlation ($p>0.05$) were excluded (gray boxes). See Supplemental Materials (Supp. Fig. 17) for an annotated version of this figure.

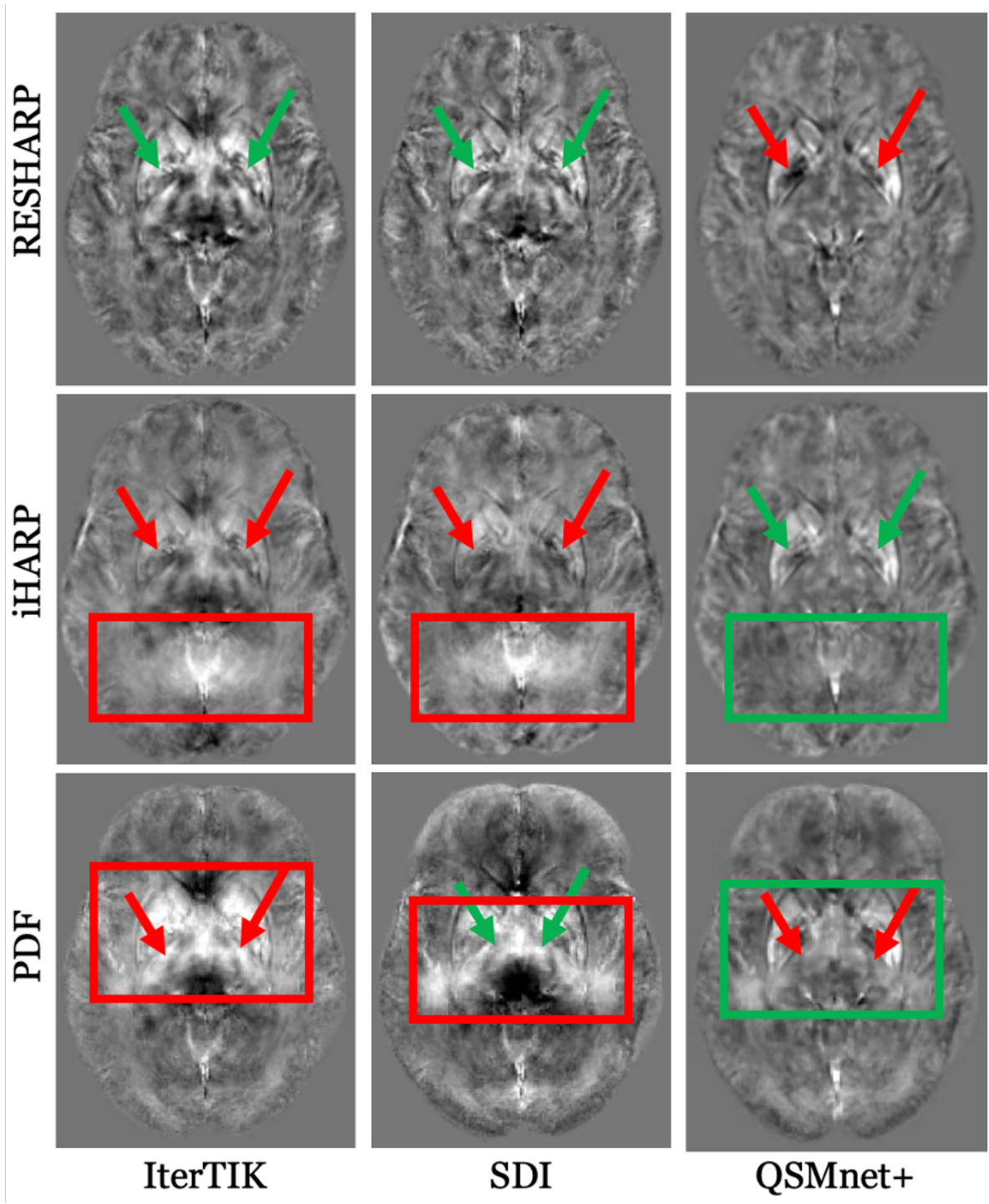

Fig. 6. Shown above are the WB-referenced group-average susceptibility difference (follow-up - baseline) maps. Y-axis shows the BFR, while x-axis displays the inversion algorithm used. Difference maps are portrayed in the same slice of the template as shown in Supp. Fig. 2, and were DGM-normalized. Color-bar was set accordingly (-0.5 to

0.6) to ensure clear visualization of regions and artifacts. Arrows point at the bi-lateral GP region within RESHARP and iHARP maps, while, thalamic region within the PDF maps, indicating the over-time susceptibility change detected, with green being consistent with increasing iron concentration, and red being inconsistent. Red boxes highlight a region affected by artifacts in iterTik and SDI. Artifacts were absent (green box) in QSMnet+.

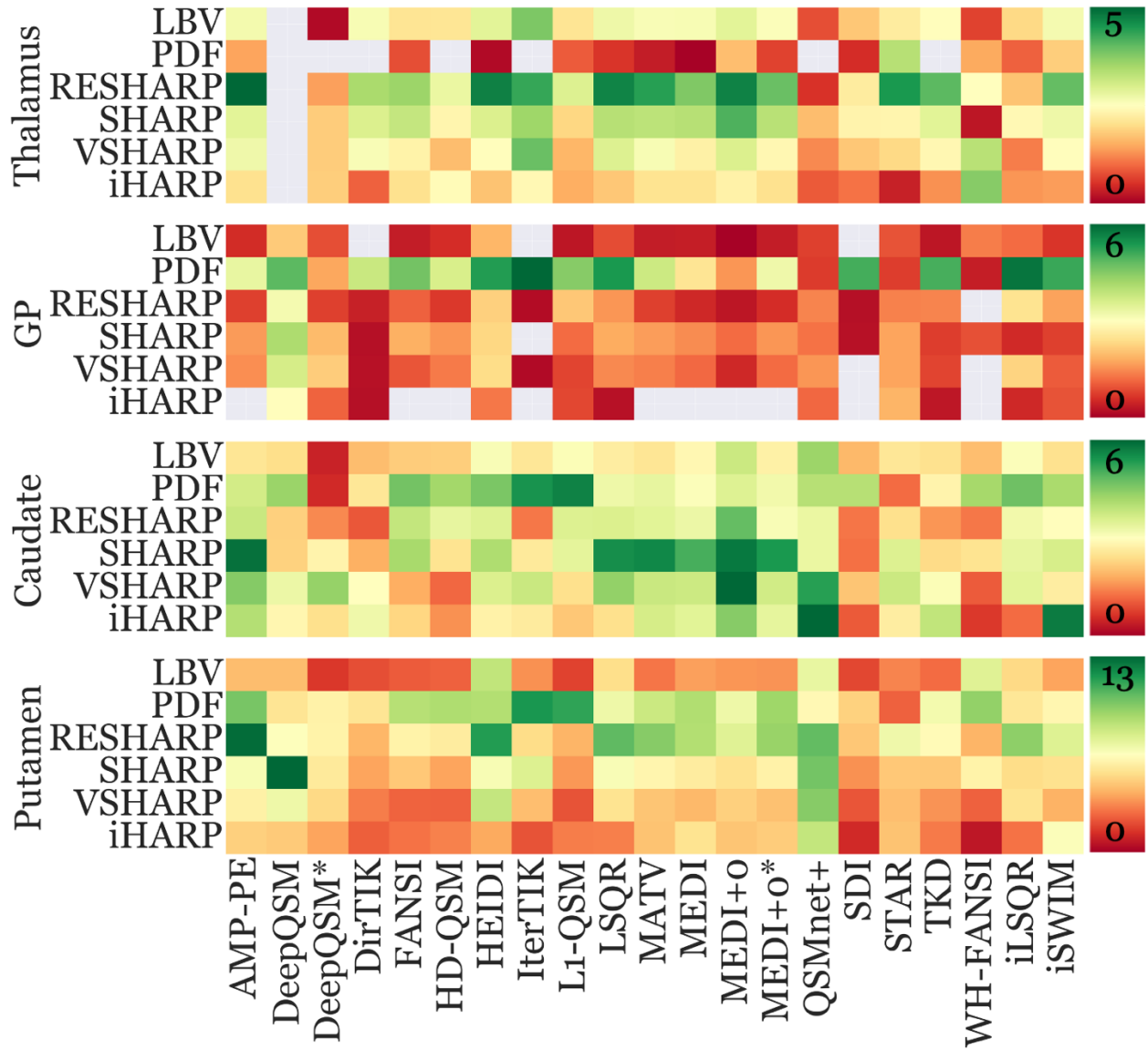

Fig. 7. Pipeline sensitivity toward aging-related susceptibility changes using WB reference. Each row corresponds to a combination of a BFR algorithm and DGM ROI. Each column represents an inversion algorithm. Susceptibility changes incompatible with H&S were excluded (gray box) to facilitate visualization. Each of the four regions (blocks of rows) has its own color bar on the right, with green indicating high sensitivity and red low. See Supplemental Materials (Supp. Fig. 20) for an annotated version of this figure.

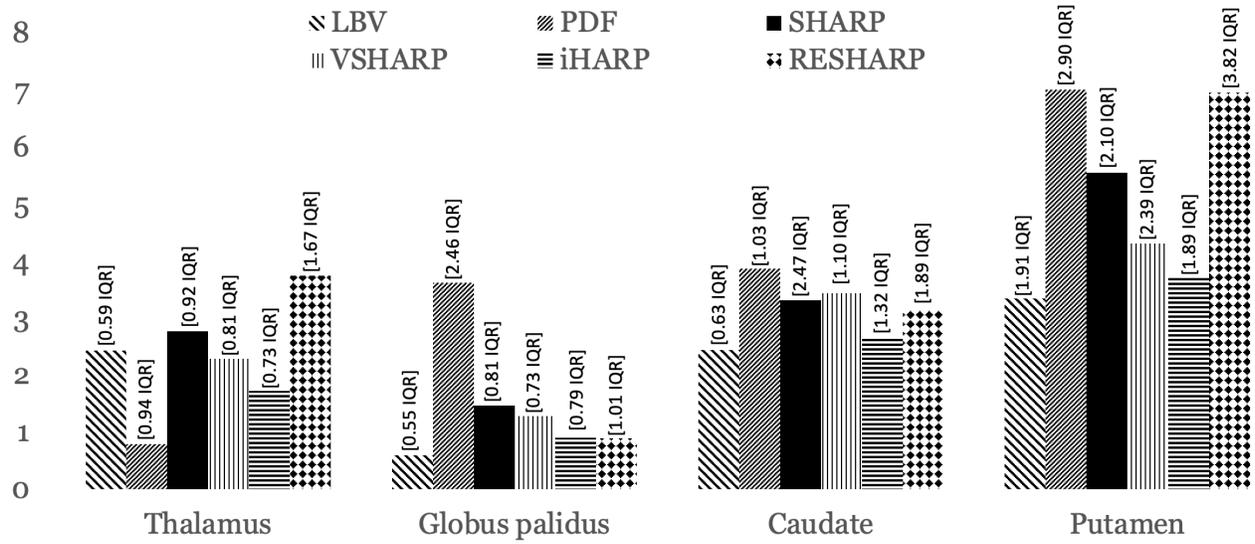

Fig. 8. WB-referenced median sensitivity of each BFR algorithm across all inversion algorithms. Y-axis portrays the median sensitivity. Inversion algorithms detecting temporal susceptibility changes inconsistent with H&S were excluded from the median.

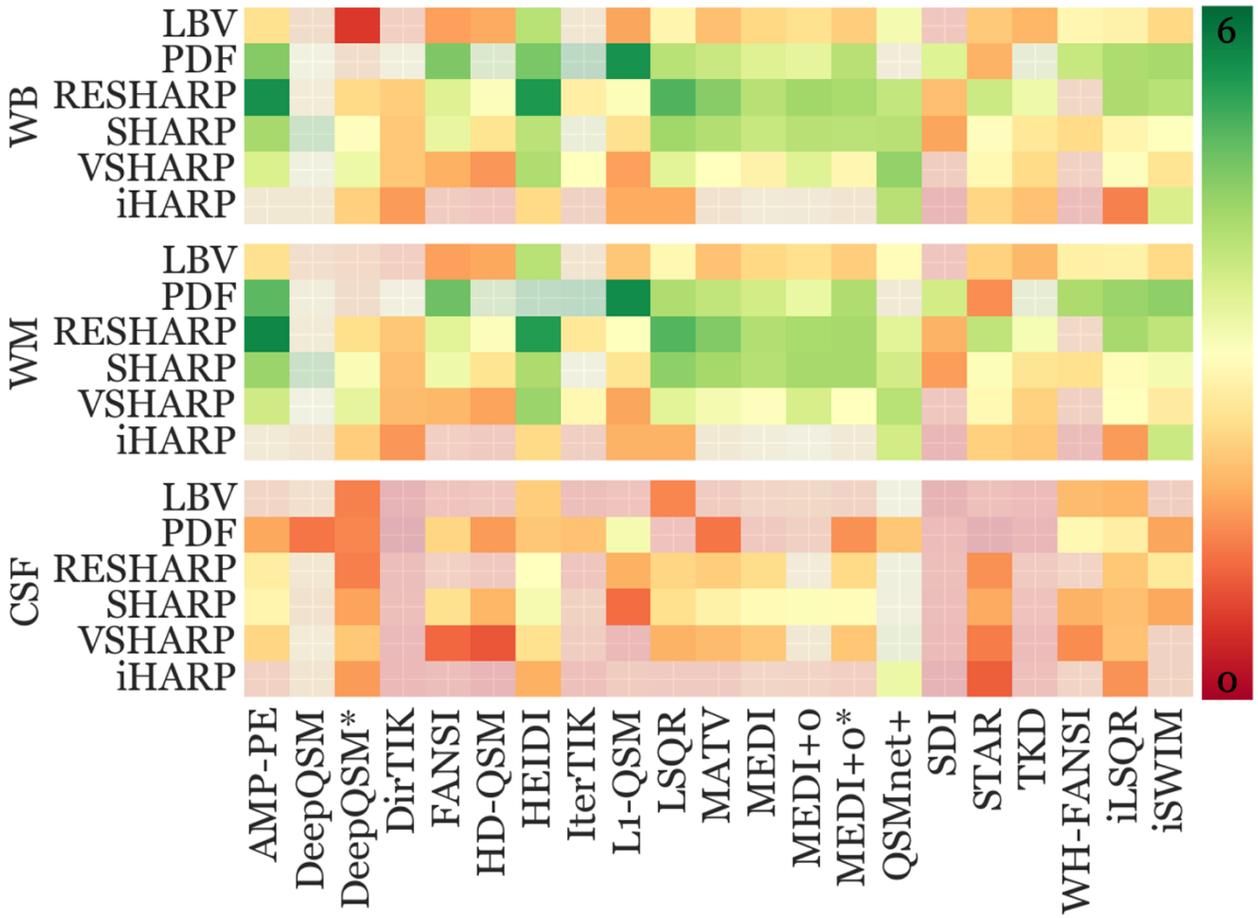

Fig. 9. Global performance sensitivity metric as defined in Eq. 7 (right). Each row corresponds to a BFR algorithm, and each column represents an inversion algorithm. Pipelines that yielded regional changes incompatible with H&S in any of the regions are translucent instead of gray boxes (distinguishable by crisscross within the boxes) in the other figures (due to numerous exclusions) to facilitate visualization. Green indicates high sensitivity, and vice versa for red. See Supplemental Materials (Supp. Fig. 23) for an annotated version of this figure.

# Supplementary Figures

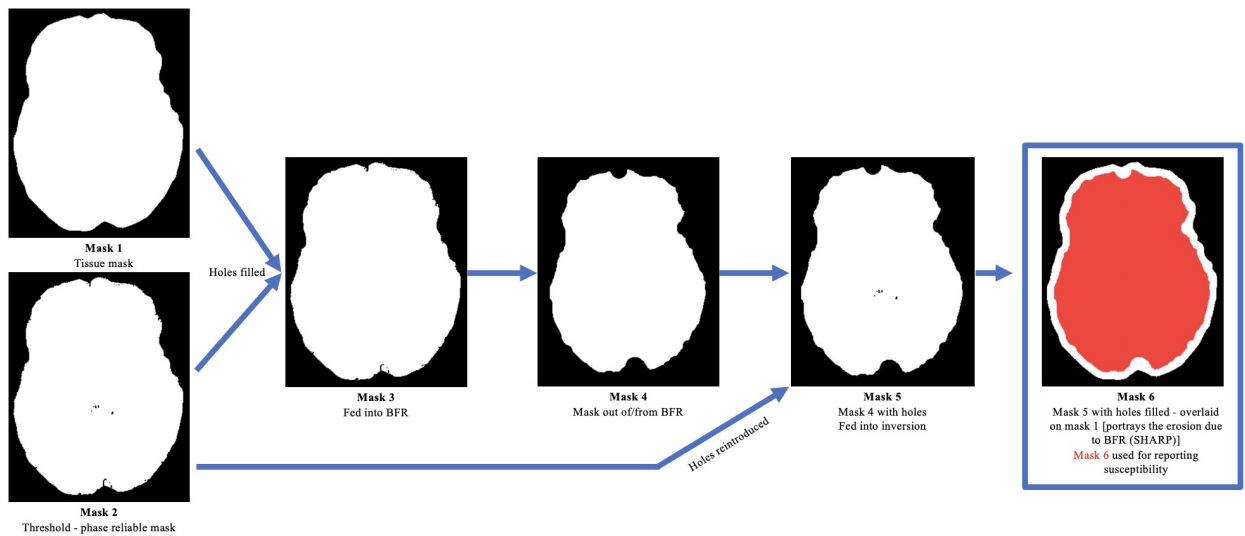

Supplementary Fig. 1 Schematic of the masking procedure utilized to generate susceptibility maps using SHARP as an exemplary BFR. Displays a mask obtained from a representative subject within the study cohort.

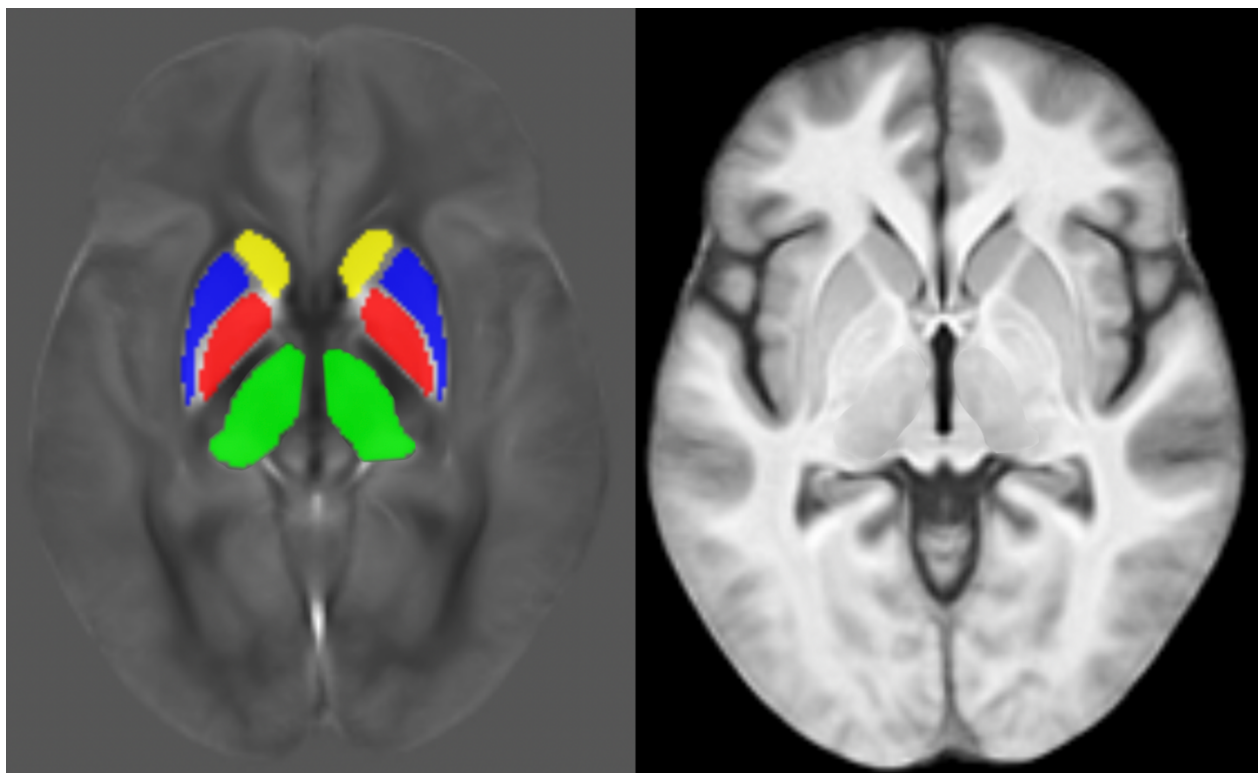

Supplementary Fig. 2 QSM and T1w contrast templates generated using the bi-parametric approach for this study, respectively. Bi-lateral ROIs shown are thalamus (green), GP (red), caudate (yellow), and putamen (blue). The contrast for QSM template was set to -0.10 to 0.20.

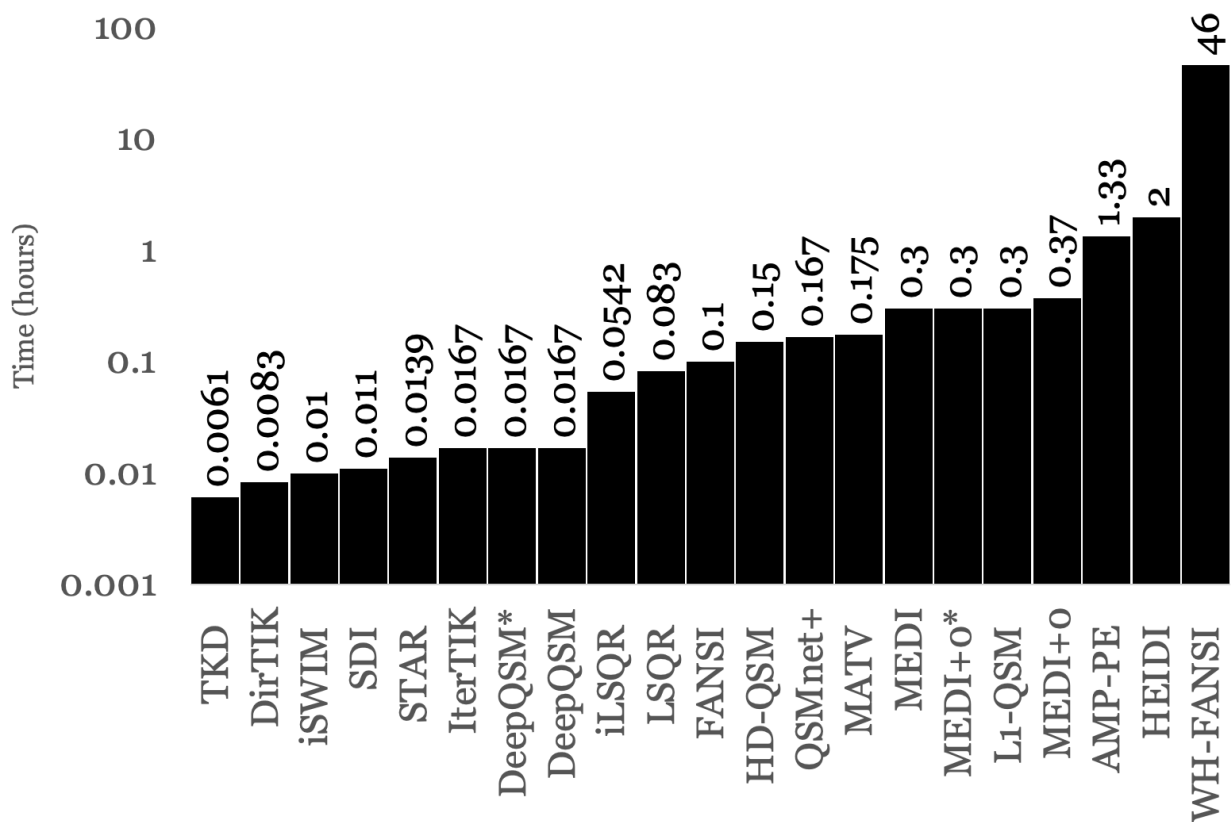

Supplementary Fig. 3 Log-scaled average dipole inversion times in ascending order (4 CPUs used).

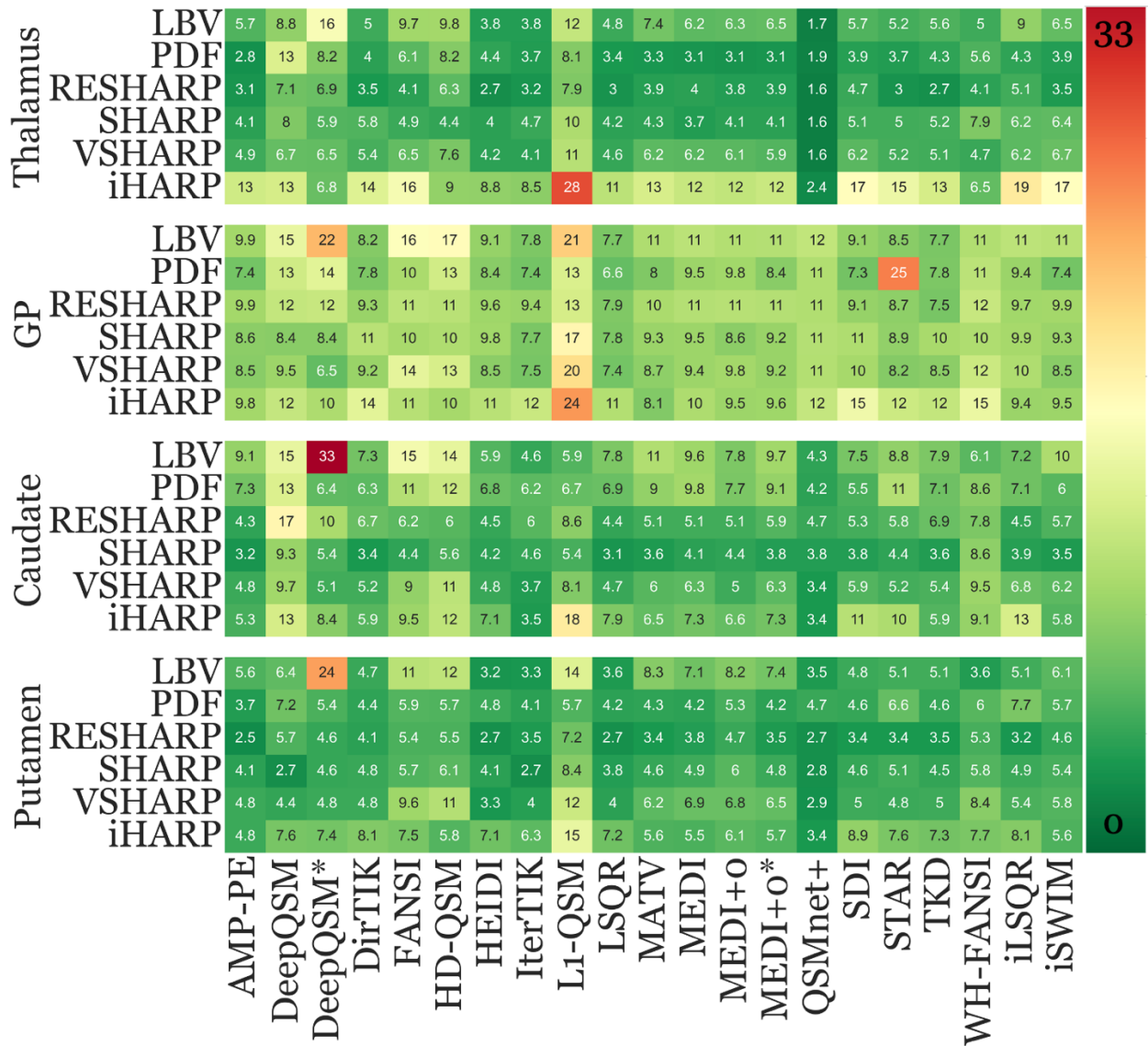

Supplementary Fig. 4 WB-referenced normalized reproducibility findings of each pipeline and region. Lower values (green) represent higher reproducibility, and vice versa for red. Each horizontal panel corresponds to one specific DGM region that is denoted at the left-hand side of the panel. Each row corresponds to a specific BFR algorithm (listed on the left-hand side) while each column represents an inversion algorithm (listed at the bottom).

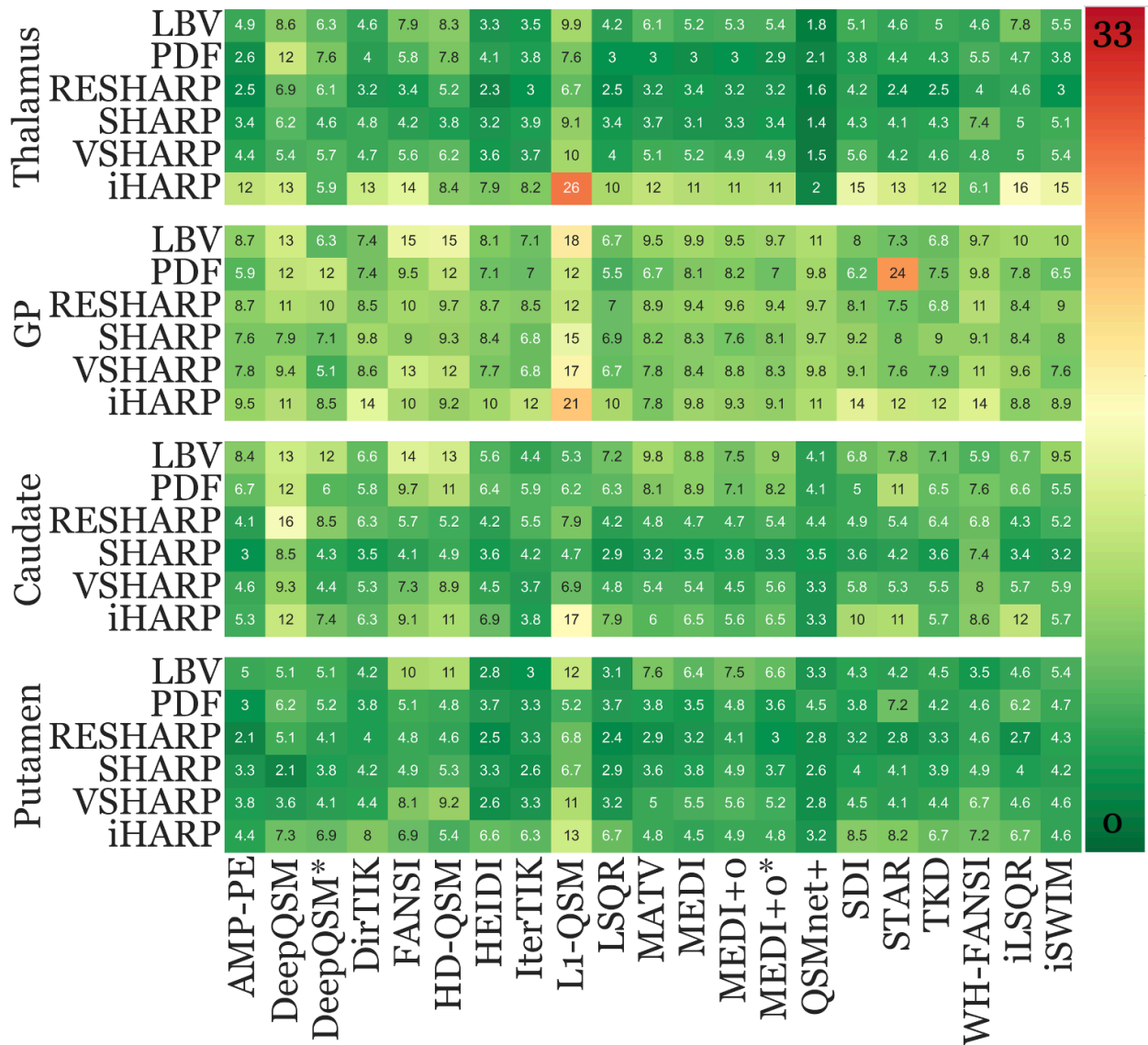

Supplementary Fig. 5 WM-referenced normalized reproducibility findings of each pipeline and region. Lower values (green) represent higher reproducibility, and vice versa for red. Each horizontal panel corresponds to one specific DGM region that is denoted at the left-hand side of the panel. Each row corresponds to a specific BFR algorithm (listed on the left-hand side) while each column represents an inversion algorithm (listed at the bottom).

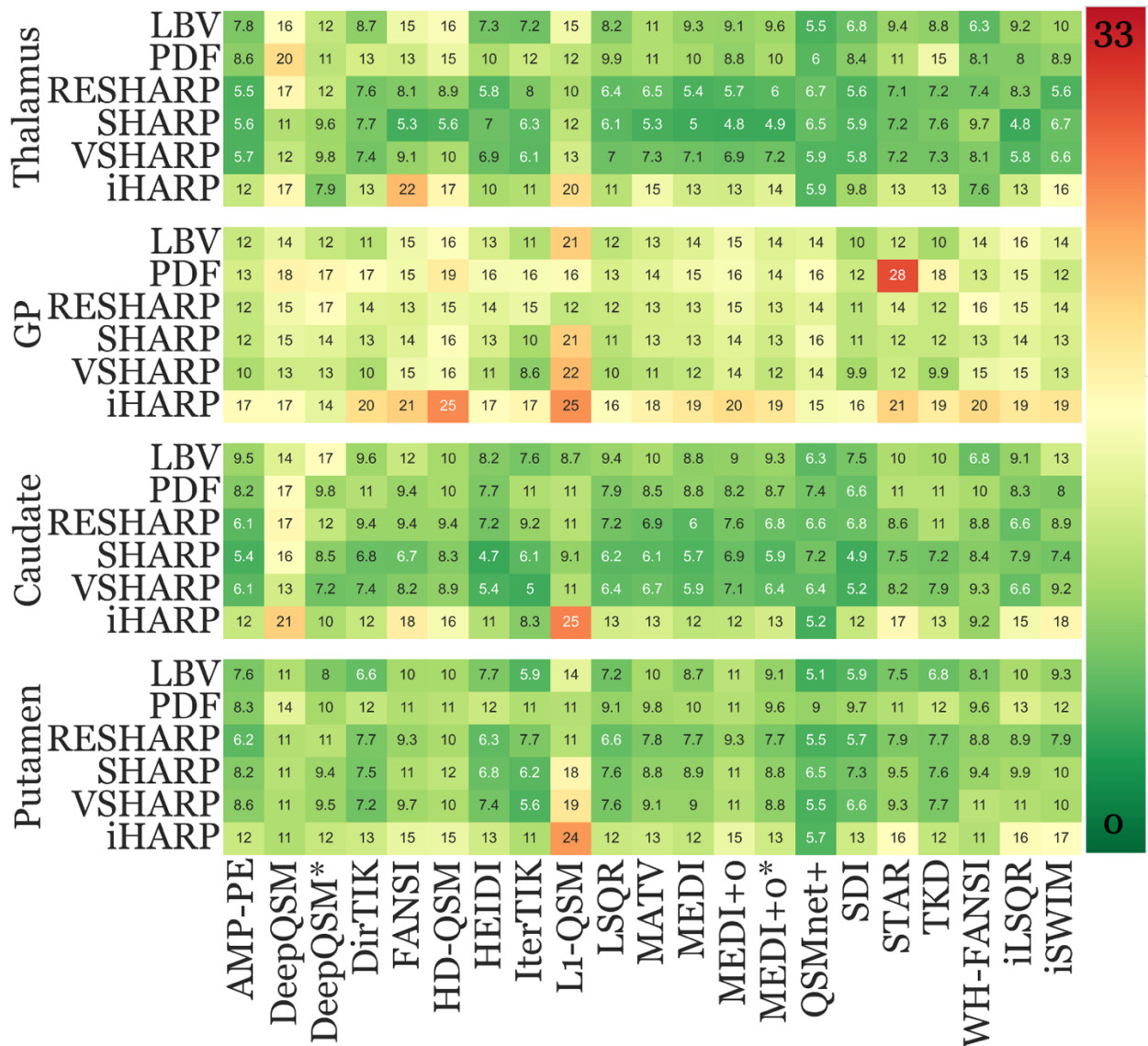

Supplementary Fig. 6 CSF-referenced normalized reproducibility findings of each pipeline and region. Lower values (green) represent higher reproducibility, and vice versa for red. Each horizontal panel corresponds to one specific DGM region that is denoted at the left-hand side of the panel. Each row corresponds to a specific BFR algorithm (listed on the left-hand side) while each column represents an inversion algorithm (listed at the bottom).

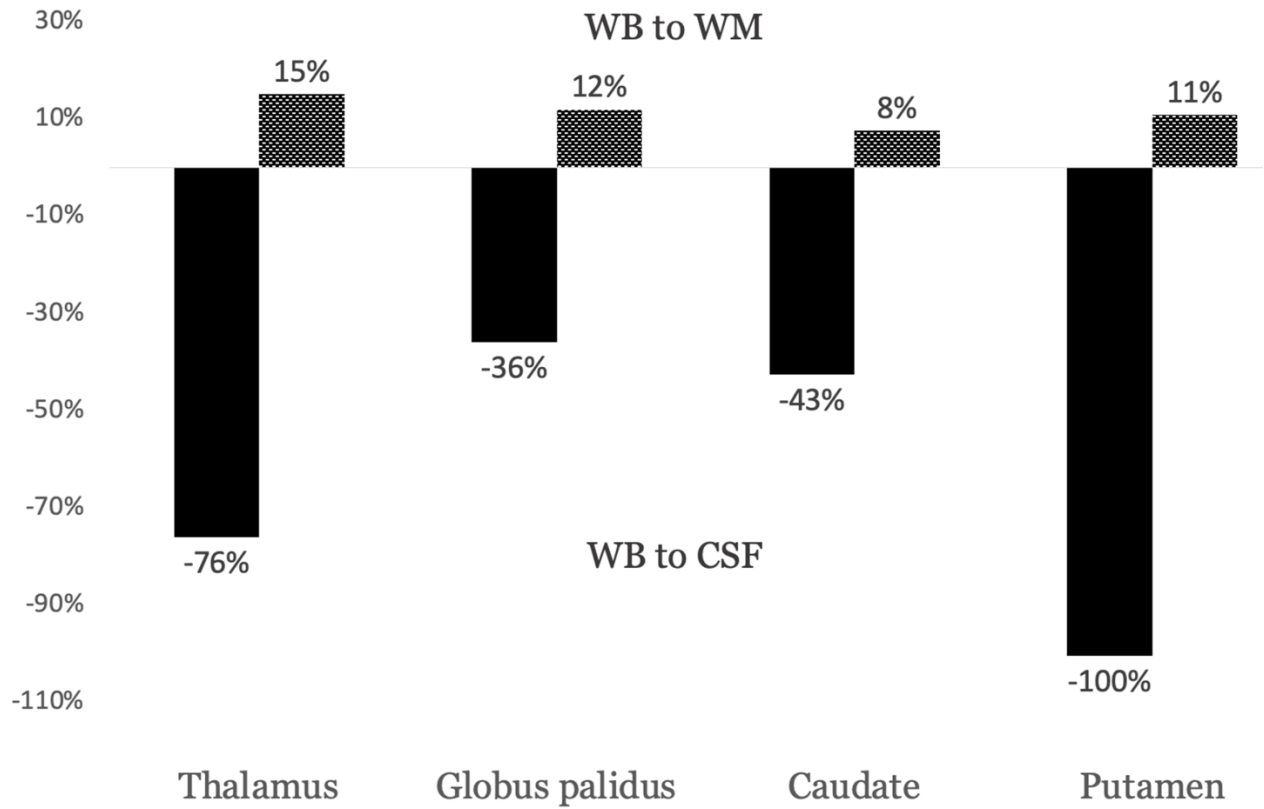

Supplementary Fig. 7 Median percentage increase/decrease (with [IQR]) in reproducibility across all inversion algorithms when the reference region was changed from WB to CSF (blue) or WM (orange). CSF and WM-referenced normalized reproducibility findings were subtracted from brain-referenced findings to compute the percentage change in reproducibility due to the switch from the WB as the reference region. Y-axis shows the reproducibility change (in %) while x-axis shows the ROI. Top part of the plot shows the change in reproducibility when reference region was switched from WB to WM, while bottom shows from WB to CSF.

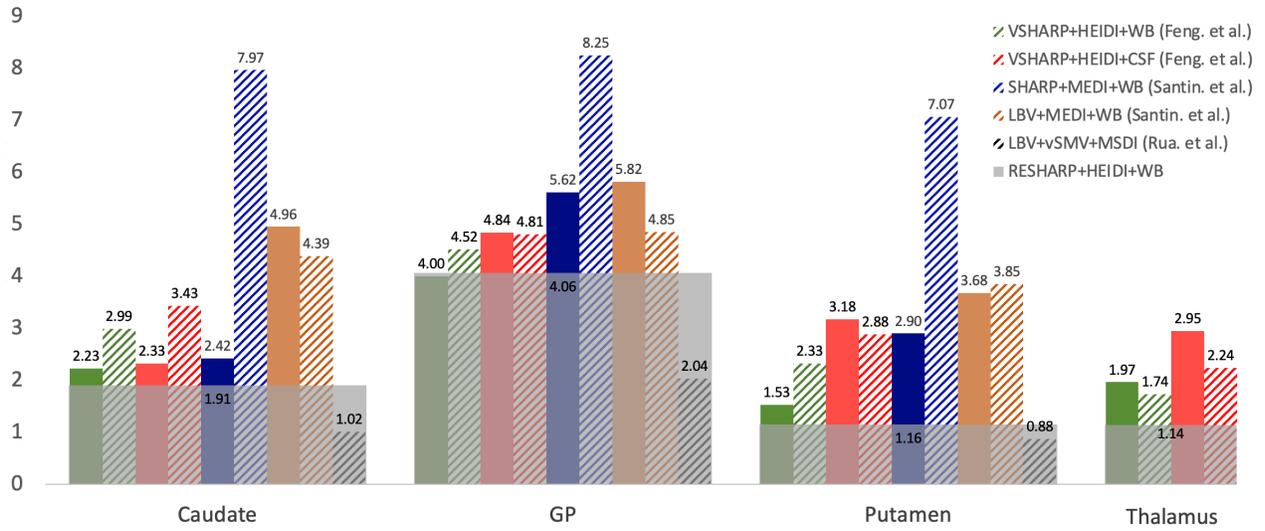

Supplementary Fig. 8 Comparison of raw scan-rescan reproducibility values (this study - solid bars) with previously reported (striped bar) pipeline reproducibility. X-axis displays the DGM region while y-axis displays the reproducibility in ppb. Reproducibility values from the present study were obtained with pipelines matched to previous studies, except for LBV+vSMV+MSDI, which was not investigated within our study. The transparent gray bar overlaid indicates one of the highest reproducibility findings across all pipelines investigated in the present study.

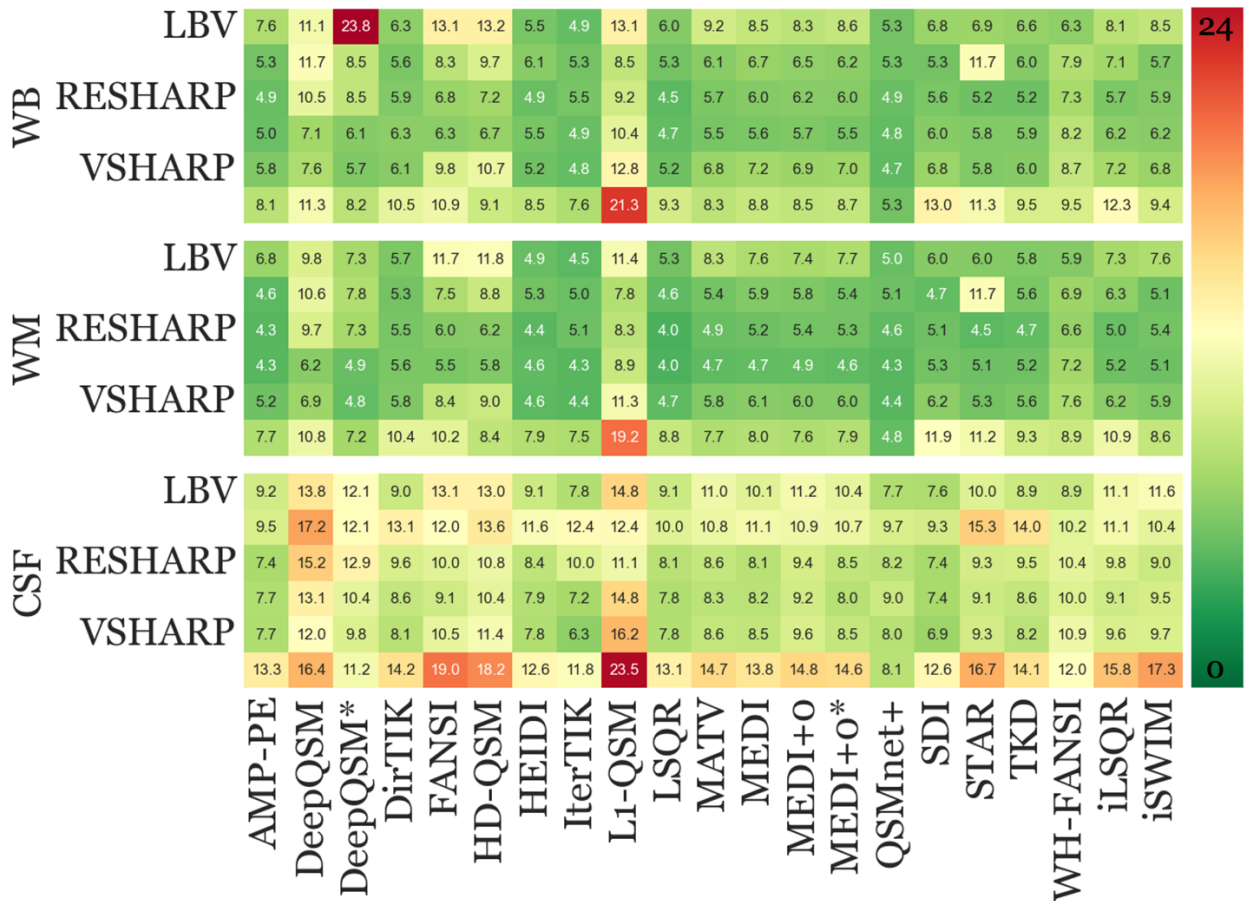

Supplementary Fig. 9 Global performance reproducibility heatmap according to Eq. 4. The color-coding and the arrangement of BFR and inversion algorithms mirrors that of supplementary Fig. 2a, 2b, and 2c. In this figure, each panel represents a specific reference region (listed on the left-hand side).

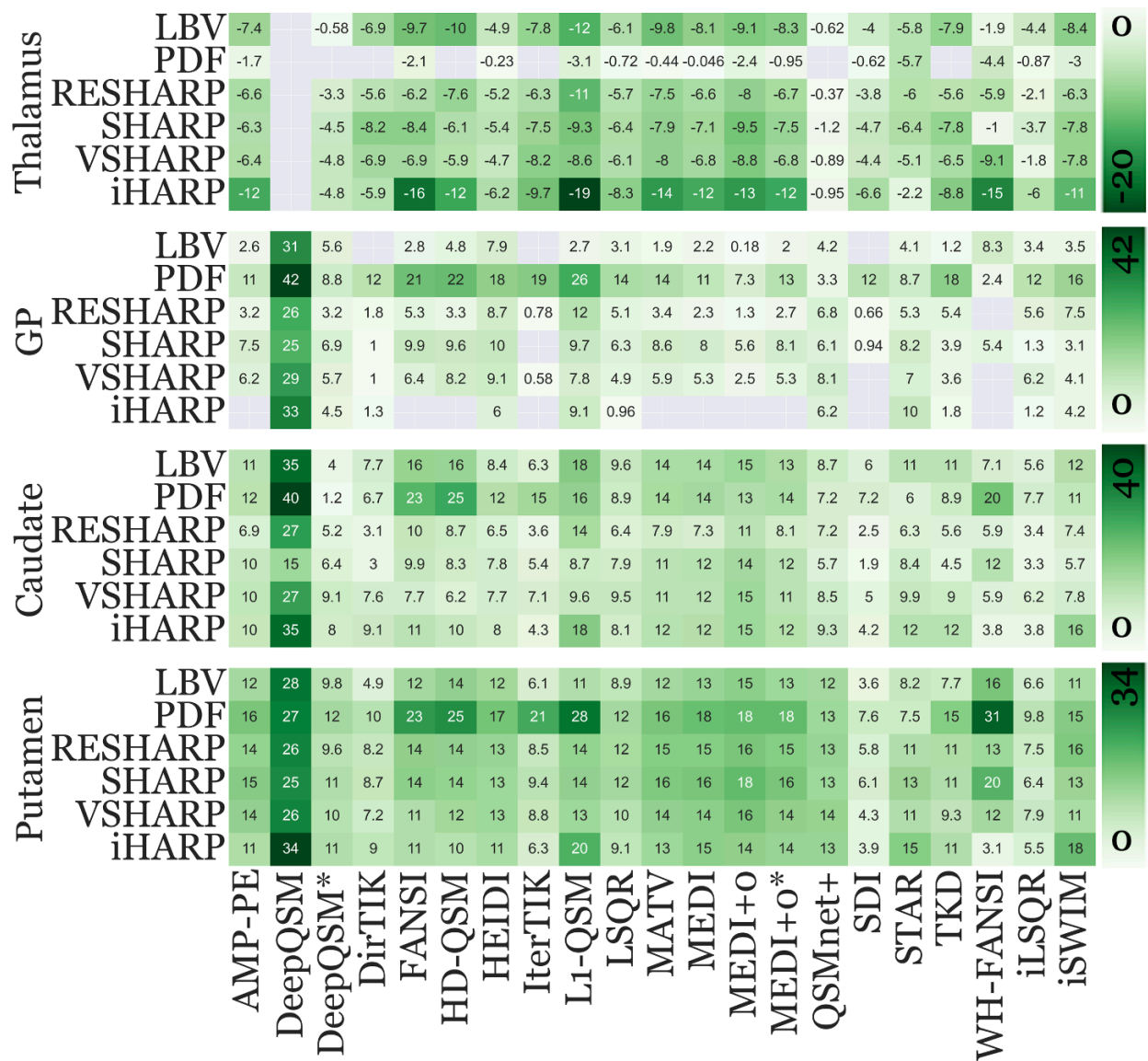

Supplementary Fig. 10 Raw WB-referenced temporal susceptibility (ppb) findings, respectively, of each pipeline. Each row corresponds to a BFR algorithm while each column represents an inversion algorithm. Each panel represents a DGM region with its title on the far left.

| Thalamus | AMP-PE | DeepQSM | DeepQSM* | DirTIK | FANSI | HD-QSM | HEIDI | IterTIK | L1-QSM | LSQR | MATV | MEDI | MEDI+o | MEDI+o* | QSMnet+ | SDI | STAR | TKD | WH-FANSI | iLSQR | iSWIM |
|---|---|---|---|---|---|---|---|---|---|---|---|---|---|---|---|---|---|---|---|---|---|
| LBV | -98 | | -21 | -81 | -124 | -127 | -67 | -91 | -144 | -85 | -122 | -101 | -114 | -103 | -11 | -61 | -72 | -87 | -22 | -107 | -106 |
| PDF | -29 | | | | -34 | | -4 | | -50 | -13 | -7 | -1 | -37 | -15 | | -13 | -88 | | -59 | -27 | -48 |
| RESHARP | -101 | | -57 | -77 | -93 | -111 | -82 | -88 | -152 | -93 | -111 | -97 | -116 | -97 | -7 | -66 | -85 | -74 | -71 | -57 | -92 |
| SHARP | -82 | | -68 | -108 | -107 | -73 | -79 | -110 | -131 | -97 | -100 | -86 | -118 | -93 | -21 | -77 | -79 | -97 | -14 | -96 | -113 |
| VSHARP | -86 | | -70 | -84 | -90 | -76 | -65 | -98 | -106 | -86 | -103 | -87 | -114 | -87 | -15 | -66 | -64 | -74 | -103 | -44 | -103 |
| iHARP | -179 | | -79 | -87 | -237 | -168 | -100 | -129 | -304 | -136 | -188 | -174 | -186 | -171 | -17 | -117 | -37 | -120 | -171 | -164 | -165 |

| GP | AMP-PE | DeepQSM | DeepQSM* | DirTIK | FANSI | HD-QSM | HEIDI | IterTIK | L1-QSM | LSQR | MATV | MEDI | MEDI+o | MEDI+o* | QSMnet+ | SDI | STAR | TKD | WH-FANSI | iLSQR | iSWIM |
|---|---|---|---|---|---|---|---|---|---|---|---|---|---|---|---|---|---|---|---|---|---|
| LBV | 35 | 237 | 208 | | 36 | 59 | 108 | | 34 | 43 | 24 | 27 | 2 | 25 | 73 | | 51 | 13 | 95 | 83 | 44 |
| PDF | 174 | 481 | 157 | 196 | 346 | 334 | 303 | 298 | 418 | 252 | 213 | 170 | 110 | 193 | 64 | 245 | 134 | 269 | 31 | 374 | 262 |
| RESHARP | 49 | 270 | 55 | 24 | 80 | 49 | 136 | 11 | 177 | 82 | 50 | 34 | 19 | 40 | 126 | 11 | 76 | 70 | | 155 | 109 |
| SHARP | 98 | 255 | 105 | 14 | 126 | 114 | 148 | | 137 | 95 | 108 | 97 | 70 | 100 | 112 | 15 | 102 | 49 | 72 | 34 | 45 |
| VSHARP | 82 | 249 | 83 | 12 | 83 | 105 | 125 | 7 | 97 | 70 | 75 | 67 | 32 | 68 | 138 | | 88 | 41 | | 150 | 54 |
| iHARP | | 268 | 74 | 20 | | | 96 | | 143 | 16 | | | | | 113 | | 166 | 24 | | 32 | 64 |

| Caudate | AMP-PE | DeepQSM | DeepQSM* | DirTIK | FANSI | HD-QSM | HEIDI | IterTIK | L1-QSM | LSQR | MATV | MEDI | MEDI+o | MEDI+o* | QSMnet+ | SDI | STAR | TKD | WH-FANSI | iLSQR | iSWIM |
|---|---|---|---|---|---|---|---|---|---|---|---|---|---|---|---|---|---|---|---|---|---|
| LBV | 148 | 268 | 148 | 90 | 205 | 193 | 115 | 73 | 221 | 133 | 169 | 171 | 187 | 165 | 153 | 90 | 141 | 122 | 81 | 136 | 158 |
| PDF | 192 | 459 | 22 | 108 | 369 | 381 | 215 | 230 | 266 | 160 | 217 | 218 | 194 | 211 | 141 | 146 | 93 | 136 | 261 | 234 | 173 |
| RESHARP | 106 | 285 | 88 | 43 | 154 | 128 | 102 | 50 | 198 | 104 | 116 | 108 | 159 | 117 | 133 | 43 | 90 | 73 | 70 | 94 | 108 |
| SHARP | 132 | 155 | 97 | 40 | 126 | 98 | 114 | 79 | 123 | 119 | 140 | 144 | 176 | 142 | 105 | 31 | 104 | 56 | 160 | 86 | 82 |
| VSHARP | 139 | 232 | 132 | 93 | 100 | 80 | 106 | 85 | 119 | 133 | 146 | 148 | 190 | 146 | 146 | 76 | 124 | 102 | 67 | 149 | 103 |
| iHARP | 151 | 282 | 132 | 135 | 161 | 136 | 129 | 58 | 277 | 132 | 165 | 176 | 210 | 170 | 170 | 76 | 198 | 165 | 44 | 105 | 243 |

| Putamen | AMP-PE | DeepQSM | DeepQSM* | DirTIK | FANSI | HD-QSM | HEIDI | IterTIK | L1-QSM | LSQR | MATV | MEDI | MEDI+o | MEDI+o* | QSMnet+ | SDI | STAR | TKD | WH-FANSI | iLSQR | iSWIM |
|---|---|---|---|---|---|---|---|---|---|---|---|---|---|---|---|---|---|---|---|---|---|
| LBV | 156 | 215 | 362 | 58 | 156 | 171 | 169 | 71 | 137 | 123 | 149 | 166 | 182 | 158 | 213 | 55 | 102 | 85 | 185 | 160 | 145 |
| PDF | 271 | 312 | 213 | 162 | 380 | 376 | 291 | 332 | 459 | 217 | 257 | 272 | 275 | 281 | 250 | 153 | 116 | 225 | 414 | 299 | 245 |
| RESHARP | 212 | 278 | 165 | 112 | 212 | 204 | 207 | 119 | 196 | 193 | 217 | 217 | 235 | 216 | 246 | 101 | 154 | 147 | 154 | 205 | 231 |
| SHARP | 201 | 262 | 168 | 114 | 185 | 164 | 188 | 138 | 198 | 180 | 200 | 200 | 226 | 200 | 235 | 101 | 161 | 135 | 264 | 167 | 188 |
| VSHARP | 184 | 230 | 146 | 87 | 145 | 158 | 172 | 105 | 155 | 141 | 180 | 183 | 212 | 181 | 234 | 65 | 134 | 106 | 136 | 190 | 149 |
| iHARP | 165 | 275 | 182 | 133 | 155 | 138 | 185 | 84 | 314 | 150 | 176 | 208 | 196 | 188 | 244 | 69 | 246 | 151 | 35 | 152 | 274 |

Supplementary Fig. 11 Normalized WB-referenced temporal susceptibility (ppb) findings, respectively, of each pipeline. Each row corresponds to a BFR algorithm while each column represents an inversion algorithm. Each panel represents a DGM region with its title on the far left.

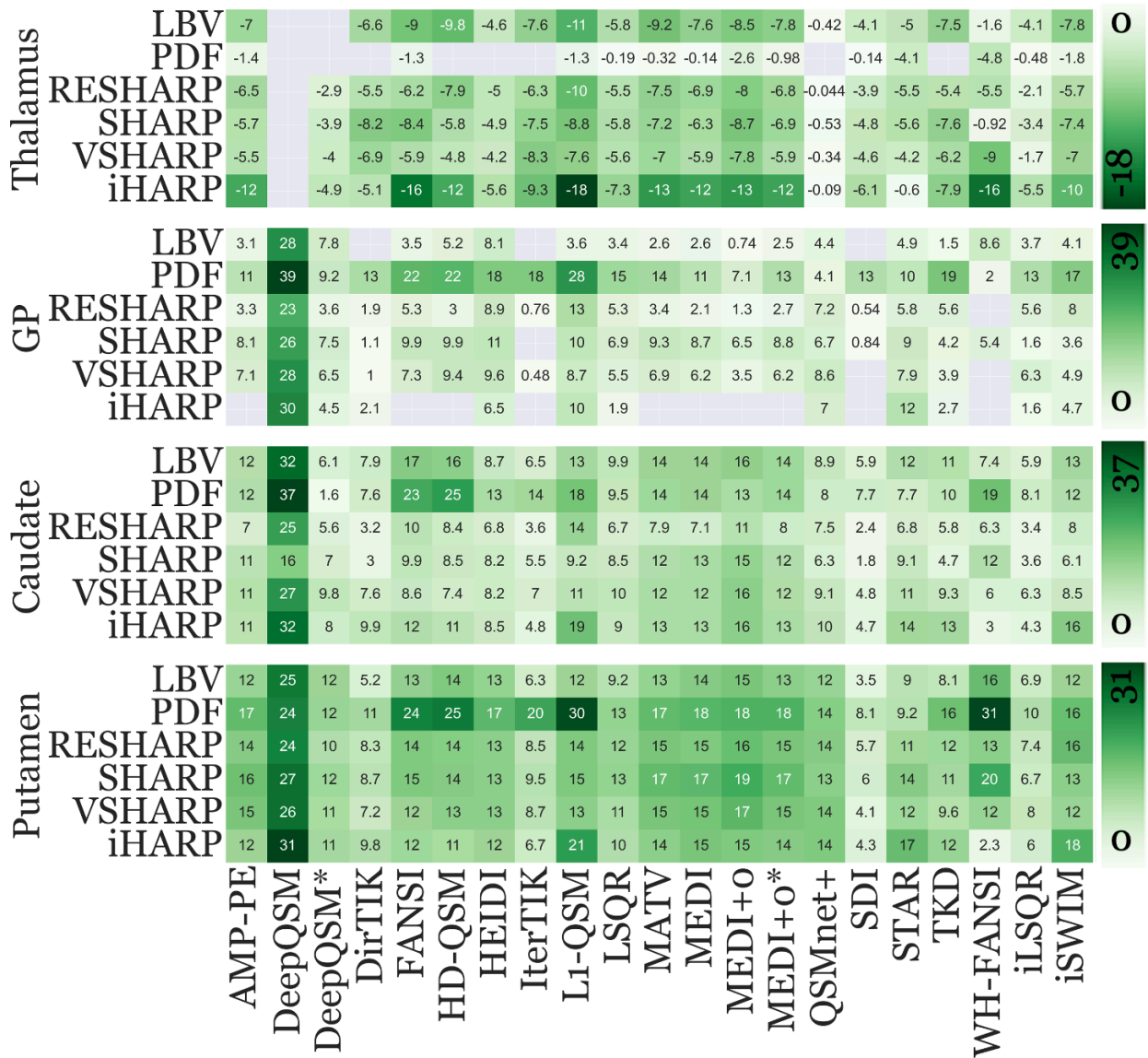

Supplementary Fig. 12 Raw WM-referenced temporal susceptibility (ppb) findings, respectively, of each pipeline. Each row corresponds to a BFR algorithm while each column represents an inversion algorithm. Each panel represents a DGM region with its title on the far left.

## Thalamus

| | AMP-PE | DeepQSM | DeepQSM* | DirTIK | FANSI | HD-QSM | HEIDI | IterTIK | L1-QSM | LSQR | MATV | MEDI | MEDI+o | MEDI+o* | QSMnet+ | SDI | STAR | TKD | WH-FANSI | iLSQR | iSWIM |
|---|---|---|---|---|---|---|---|---|---|---|---|---|---|---|---|---|---|---|---|---|---|
| LBV | -85 | | | -71 | -107 | -112 | -59 | | -82 | -130 | -75 | -105 | -88 | -99 | -90 | -7 | -58 | -57 | -76 | -17 | -93 | -91 |
| PDF | -21 | | | -19 | | | -19 | -3 | -5 | -2 | -36 | -14 | | -3 | -57 | | -60 | -14 | -27 | | | |
| RESHARP | -91 | | -47 | -70 | -85 | -106 | -72 | -82 | -133 | -82 | -101 | -92 | -106 | -90 | -1 | -62 | -72 | -65 | -62 | -53 | -77 |
| SHARP | -67 | | -52 | -95 | -96 | -63 | -65 | -96 | -110 | -78 | -81 | -69 | -97 | -76 | -9 | -69 | -63 | -83 | -11 | -78 | -94 |
| VSHARP | -67 | | -53 | -75 | -70 | -56 | -52 | -89 | -86 | -71 | -80 | -68 | -91 | -68 | -5 | -62 | -47 | -63 | -93 | -38 | -83 |
| iHARP | -165 | | -74 | -72 | -218 | -154 | -87 | -116 | -275 | -114 | -171 | -162 | -171 | -158 | -1 | -102 | -10 | -102 | -171 | -145 | -149 |

Range: 0 to -275

## GP

| | | | | | | | | | | | | | | | | | | | | | | |
|---|---|---|---|---|---|---|---|---|---|---|---|---|---|---|---|---|---|---|---|---|---|---|
| LBV | 37 | 205 | 104 | | 41 | 60 | 104 | | 43 | 44 | 29 | 30 | 9 | 28 | 70 | | 55 | 16 | 92 | 83 | 47 |
| PDF | 165 | 415 | 149 | 191 | 331 | 310 | 282 | 257 | 411 | 239 | 197 | 155 | 98 | 177 | 73 | 232 | 145 | 260 | 24 | 354 | 258 |
| RESHARP | 46 | 231 | 57 | 24 | 73 | 41 | 129 | 10 | 170 | 79 | 46 | 28 | 17 | 35 | 120 | 9 | 76 | 68 | | 143 | 108 |
| SHARP | 95 | 237 | 100 | 12 | 113 | 106 | 139 | | 128 | 92 | 104 | 95 | 71 | 97 | 110 | 12 | 100 | 46 | 66 | 38 | 45 |
| VSHARP | 85 | 230 | 84 | 11 | 87 | 109 | 120 | 5 | 98 | 69 | 80 | 71 | 41 | 72 | 133 | | 90 | 40 | | 138 | 57 |
| iHARP | | 239 | 68 | 29 | | | 100 | | 151 | 30 | | | | | 116 | | 187 | 35 | | 43 | 69 |

Range: 0 to 415

## Caudate

| | | | | | | | | | | | | | | | | | | | | | | |
|---|---|---|---|---|---|---|---|---|---|---|---|---|---|---|---|---|---|---|---|---|---|---|
| LBV | 142 | 235 | 83 | 86 | 197 | 183 | 110 | 70 | 158 | 127 | 163 | 163 | 179 | 157 | 143 | 81 | 138 | 116 | 79 | 132 | 152 |
| PDF | 182 | 395 | 26 | 111 | 353 | 353 | 202 | 195 | 271 | 155 | 200 | 199 | 176 | 193 | 143 | 142 | 108 | 139 | 236 | 226 | 176 |
| RESHARP | 98 | 244 | 88 | 41 | 141 | 113 | 98 | 46 | 190 | 100 | 106 | 95 | 146 | 106 | 126 | 38 | 89 | 70 | 72 | 87 | 107 |
| SHARP | 126 | 148 | 93 | 35 | 112 | 92 | 108 | 71 | 115 | 113 | 133 | 137 | 167 | 134 | 104 | 26 | 102 | 52 | 144 | 84 | 77 |
| VSHARP | 138 | 214 | 128 | 82 | 102 | 86 | 102 | 75 | 118 | 127 | 144 | 144 | 184 | 142 | 140 | 64 | 122 | 94 | 62 | 136 | 100 |
| iHARP | 156 | 252 | 123 | 139 | 171 | 142 | 132 | 60 | 280 | 142 | 171 | 172 | 208 | 168 | 168 | 78 | 218 | 170 | 33 | 113 | 239 |

Range: 0 to 395

## Putamen

| | | | | | | | | | | | | | | | | | | | | | | |
|---|---|---|---|---|---|---|---|---|---|---|---|---|---|---|---|---|---|---|---|---|---|---|
| LBV | 150 | 184 | 160 | 56 | 152 | 163 | 160 | 68 | 143 | 118 | 144 | 158 | 175 | 152 | 197 | 49 | 102 | 82 | 177 | 155 | 140 |
| PDF | 254 | 258 | 200 | 160 | 363 | 349 | 272 | 289 | 449 | 207 | 237 | 249 | 250 | 258 | 243 | 148 | 129 | 221 | 377 | 285 | 242 |
| RESHARP | 195 | 238 | 158 | 105 | 195 | 182 | 194 | 110 | 189 | 181 | 199 | 196 | 216 | 197 | 230 | 90 | 148 | 139 | 151 | 189 | 221 |
| SHARP | 188 | 243 | 155 | 101 | 165 | 151 | 174 | 122 | 181 | 167 | 186 | 187 | 212 | 186 | 220 | 87 | 154 | 122 | 238 | 157 | 171 |
| VSHARP | 179 | 212 | 141 | 78 | 143 | 157 | 163 | 93 | 151 | 134 | 174 | 176 | 203 | 173 | 219 | 55 | 131 | 97 | 126 | 173 | 142 |
| iHARP | 169 | 246 | 170 | 137 | 165 | 143 | 184 | 84 | 315 | 159 | 181 | 202 | 195 | 186 | 234 | 72 | 265 | 157 | 25 | 158 | 269 |

Range: 0 to 449

Supplementary Fig. 13 Normalized WM-referenced temporal susceptibility (ppb) findings, respectively, of each pipeline. Each row corresponds to a BFR algorithm while each column represents an inversion algorithm. Each panel represents a DGM region with its title on the far left.

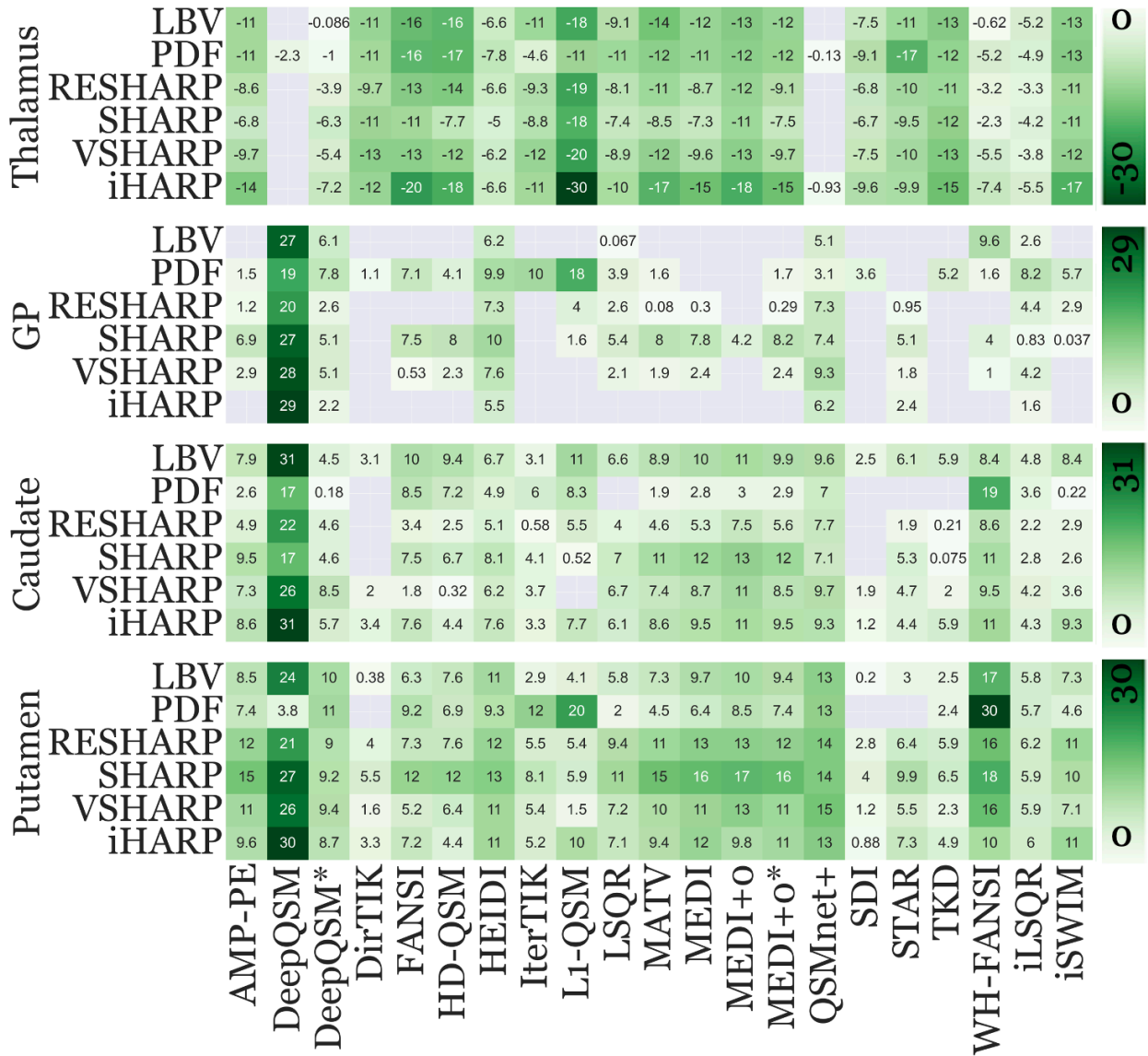

Supplementary Fig. 14 Raw CSF-referenced temporal susceptibility (ppb) findings, respectively, of each pipeline. Each row corresponds to a BFR algorithm while each column represents an inversion algorithm. Each panel represents a DGM region with its title on the far left.

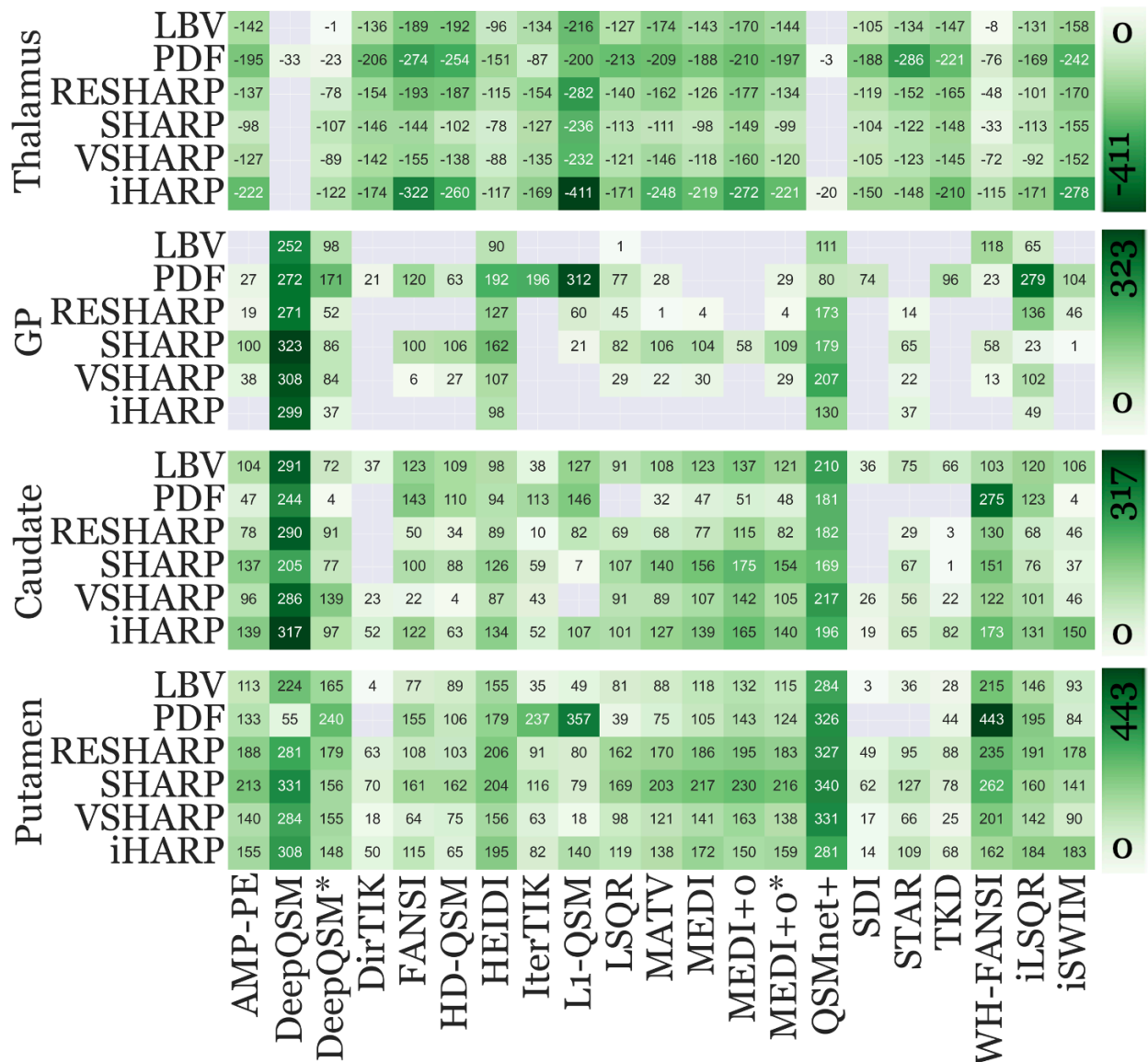

Supplementary Fig. 15 Normalized CSF-referenced temporal susceptibility (ppb) findings, respectively, of each pipeline. Each row corresponds to a BFR algorithm while each column represents an inversion algorithm. Each panel represents a DGM region with its title on the far left.

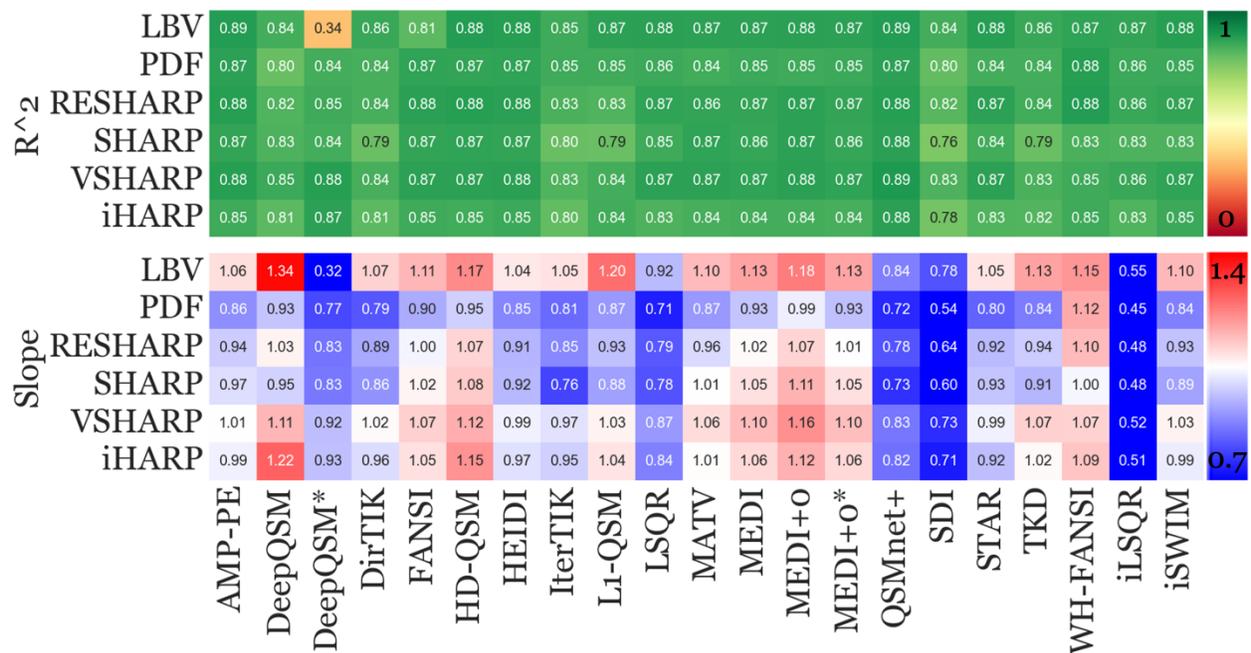

Supplementary Fig. 16. Correlation between observed DGM susceptibility and putative iron at baseline timepoint. The $R^2$ correlations are displayed at the top, while the slope is depicted at the bottom. Select pipeline plots can be viewed in Supplementary Fig. 18.

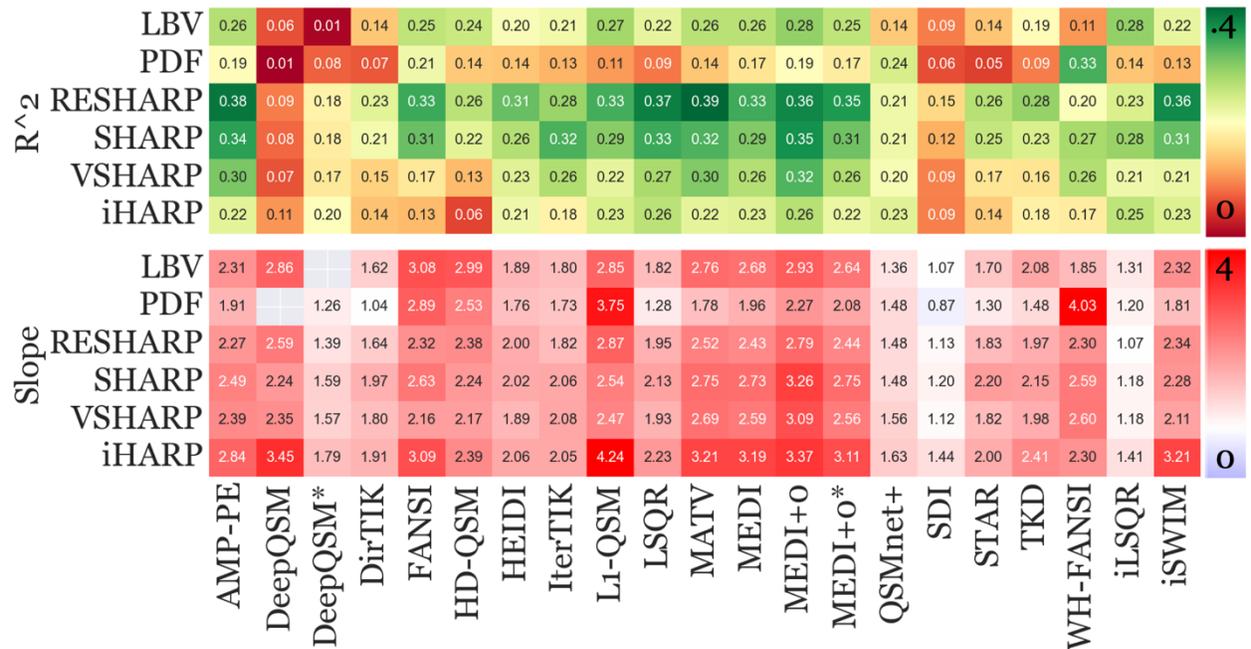

Supplementary Fig. 17. Correlation between observed DGM over-time susceptibility changes and putative iron changes. The $R^2$ correlations are displayed at the top, while the slope is depicted at the bottom. Pipelines within the slope plot that exhibited non-significant correlation ($p>0.05$) were excluded (gray boxes). Select pipeline plots can be viewed in Supplementary Fig. 19.

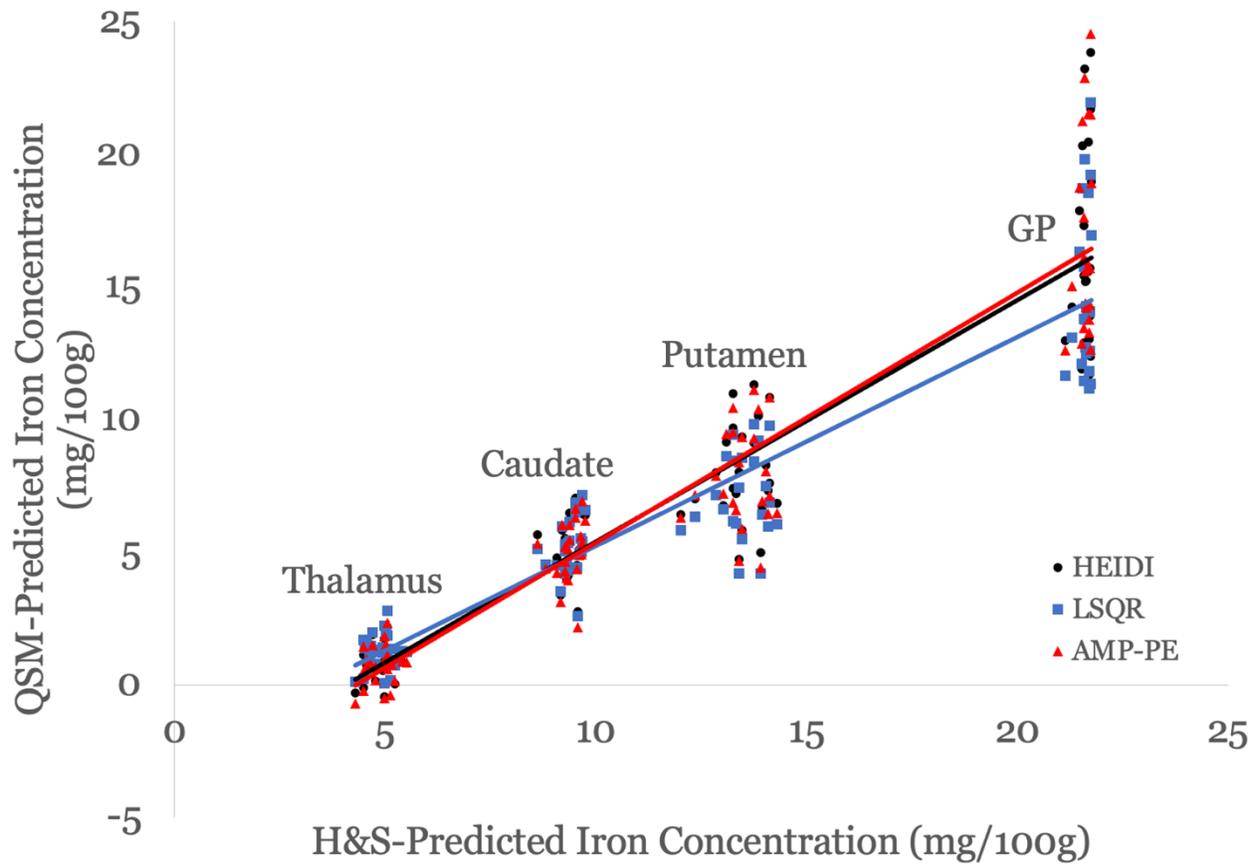

Supplementary Fig. 18. Correlation between observed DGM susceptibility and putative iron at baseline timepoint using pipelines in the 95th percentile of the WB-referenced sensitivity metric (Supp. Fig. 21 – Top panel; RESHARP+AMP-PE, HEIDI, and LSQR). Their $R^2$ correlations and slopes can be viewed in Supp. Fig 16. DGM regions are displayed on top of their data points.

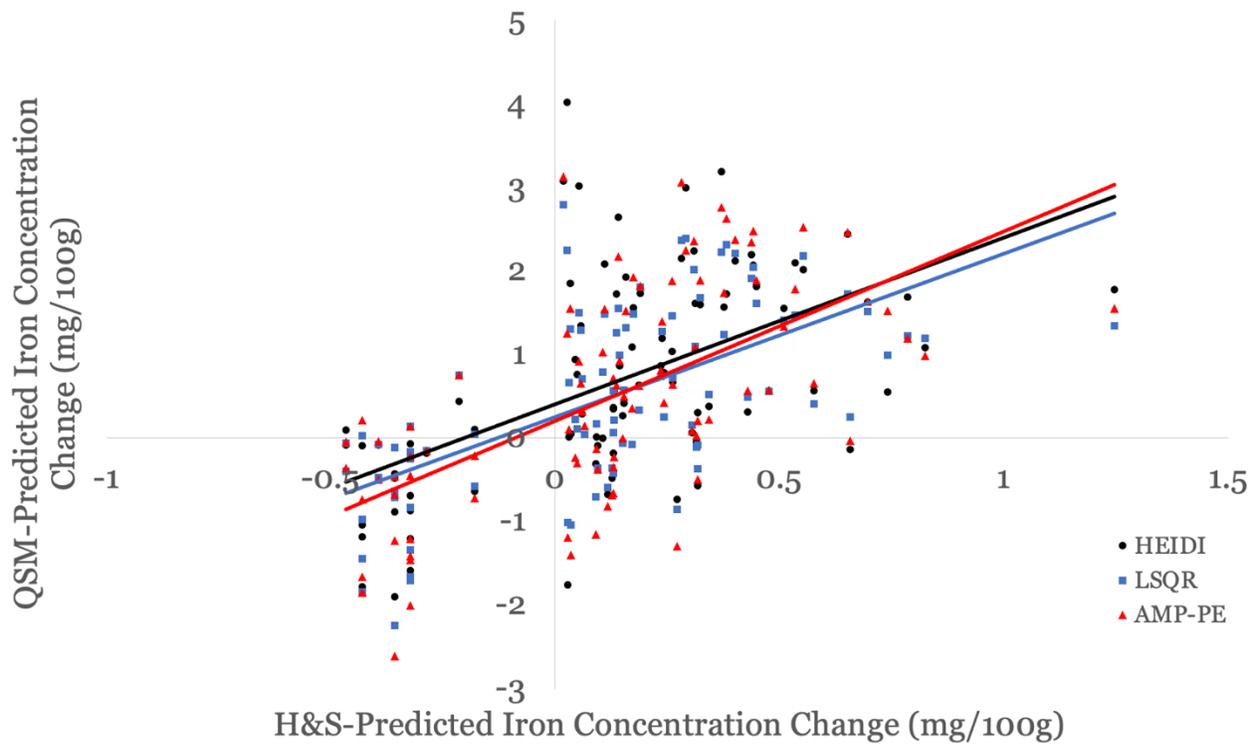

Supplementary Fig. 19. Correlation between observed DGM susceptibility and putative iron changes over-time using pipelines in the 95th percentile of the WB-referenced sensitivity metric (Supp. Fig. 21 – Top panel; RESHARP+AMP-PE, HEIDI, and LSQR). Their $R^2$ correlations and slopes can be viewed in Supp. Fig 17.

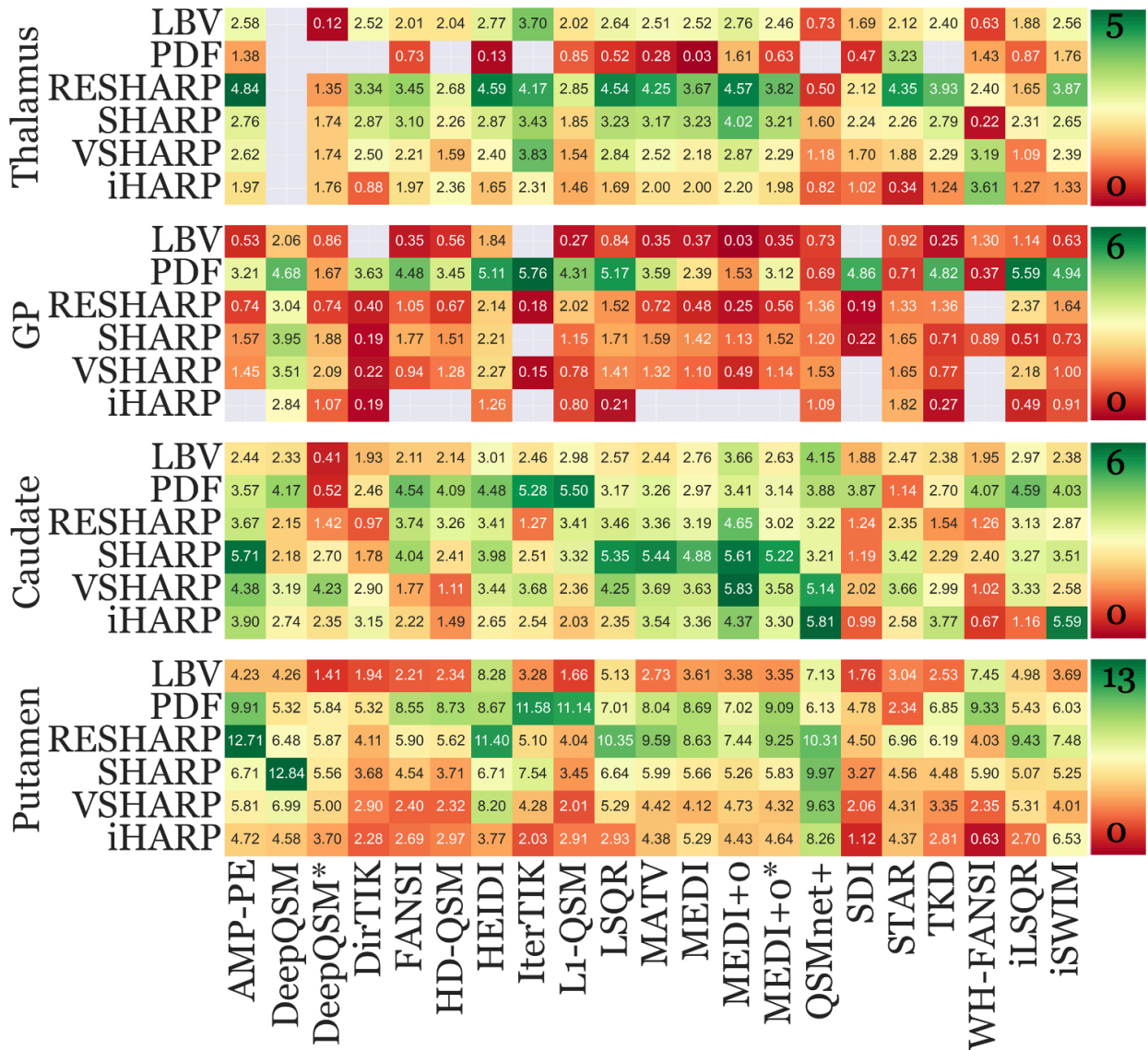

Supplementary Fig. 20 Pipeline sensitivity toward aging-related susceptibility changes using WB reference. Each row corresponds to a combination of a BFR algorithm and DGM region of interest (ROI). Each column represents an inversion algorithm. Susceptibility changes incompatible with H&S were excluded (gray box) to facilitate visualization. Each of the four regions (blocks of rows) has its own color bar on the right, with green indicating high sensitivity and red low.

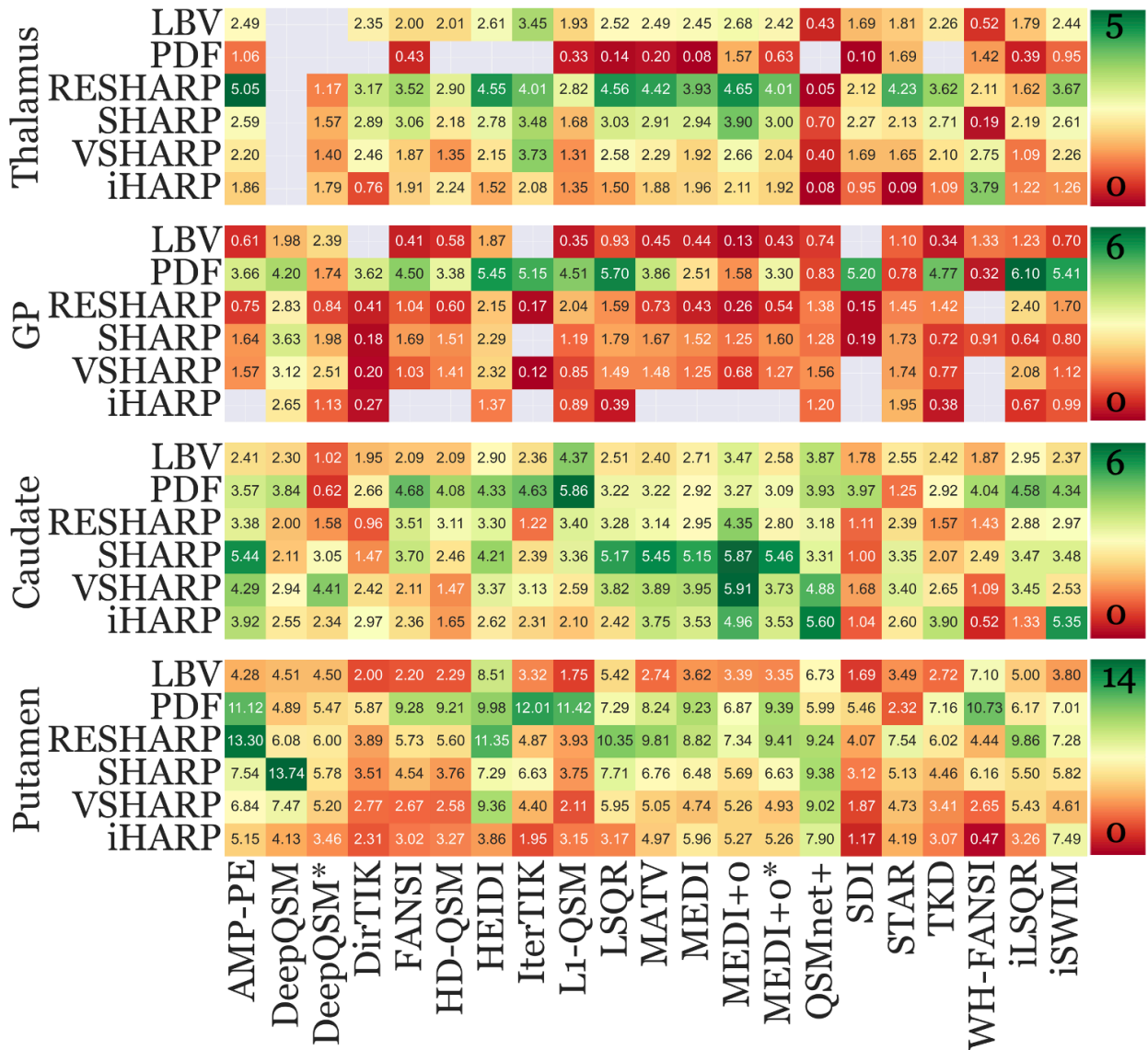

Supplementary Fig. 21 Pipeline sensitivity toward aging-related susceptibility changes using WM reference. Each row corresponds to a combination of a BFR algorithm and DGM region of interest (ROI). Each column represents an inversion algorithm. Susceptibility changes incompatible with H&S were excluded (gray box) to facilitate visualization. Each of the four regions (blocks of rows) has its own color bar on the right, with green indicating high sensitivity and red low.

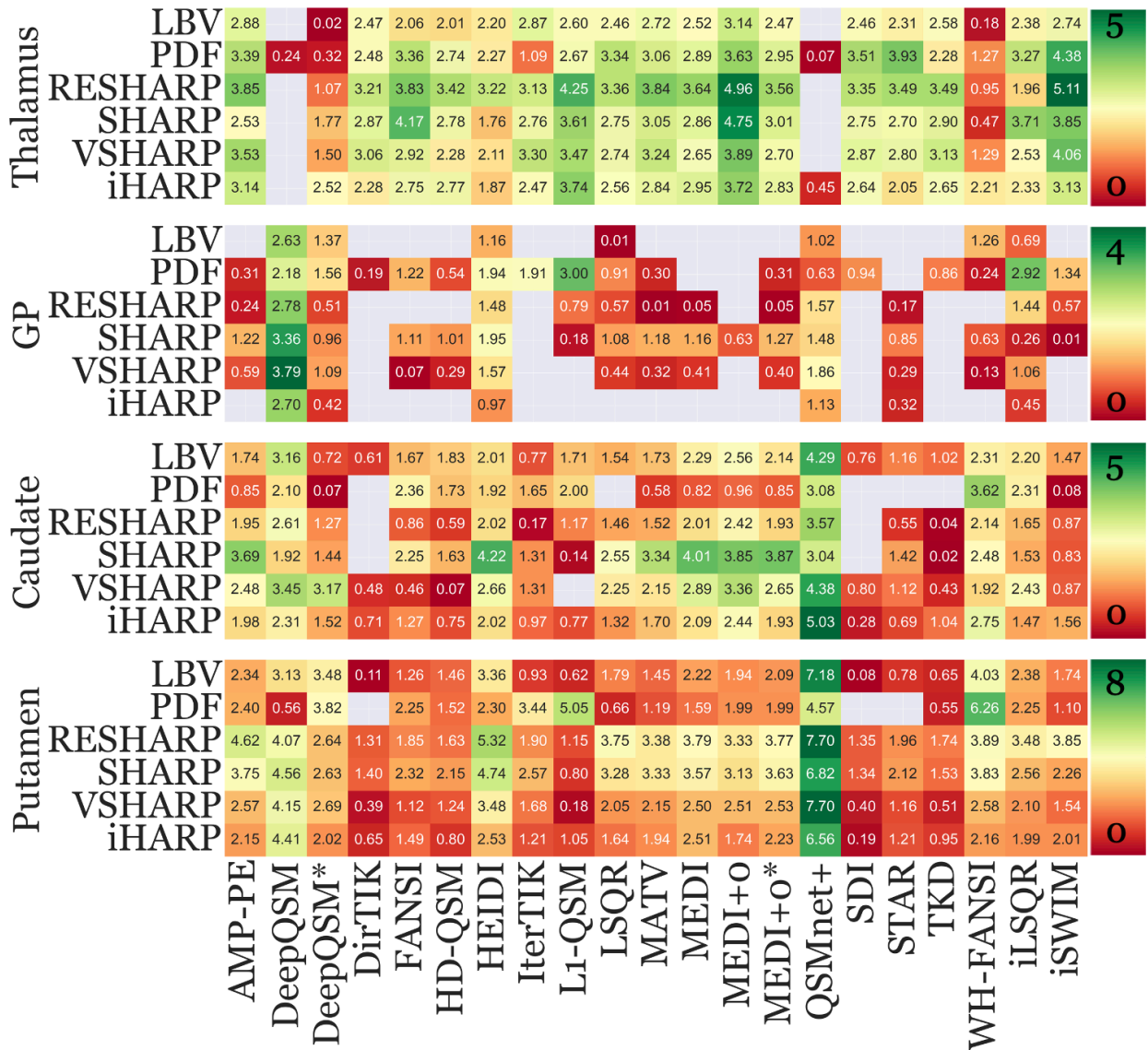

Supplementary Fig. 22 Pipeline sensitivity toward aging-related susceptibility changes using CSF reference. Each row corresponds to a combination of a BFR algorithm and DGM region of interest (ROI). Each column represents an inversion algorithm. Susceptibility changes incompatible with H&S were excluded (gray box) to facilitate visualization. Each of the four regions (blocks of rows) has its own color bar on the right, with green indicating high sensitivity and red low.

Supplementary Fig. 23 Global performance as defined in Eq. 7. Each row corresponds to a BFR algorithm, and each column represents an inversion algorithm. Pipelines that yielded regional changes incompatible with H&S in any of the regions are translucent instead of gray boxes (distinguishable by criss-cross within the boxes) in the other figures (due to numerous exclusions) to facilitate visualization. Green indicates high sensitivity, and vice versa for red.

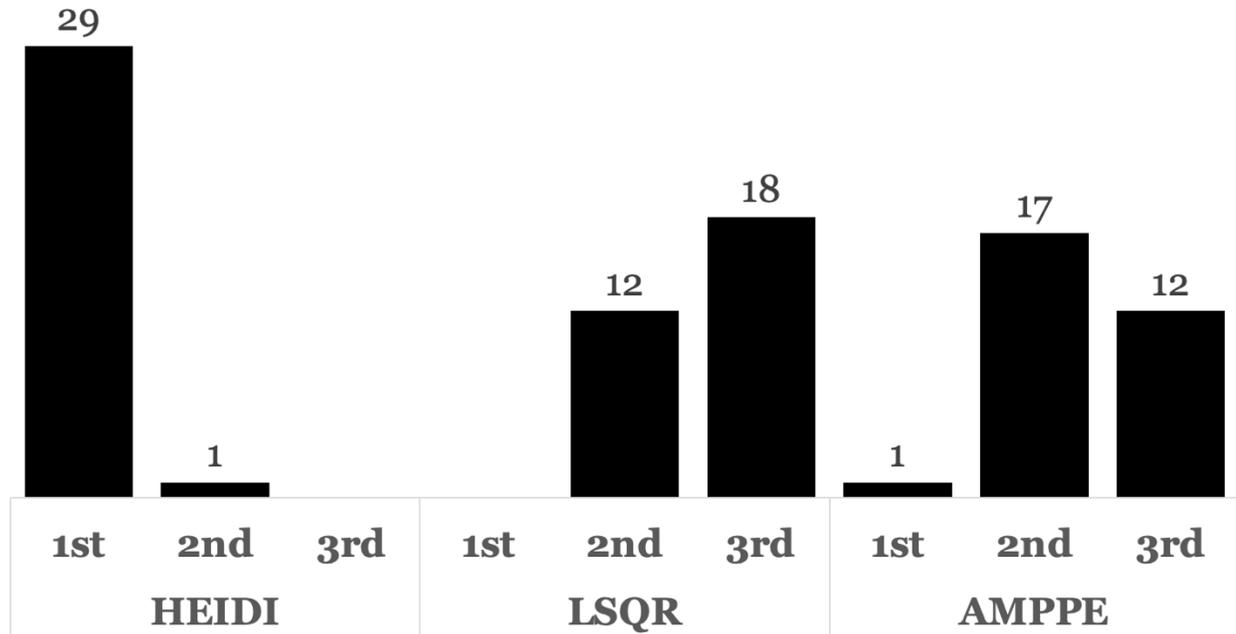

Supplementary Fig. 24 Ranking of 95th sensitivity percentile pipelines over all subjects and raters. The maximum count was 30 (3 raters x 10 subjects).

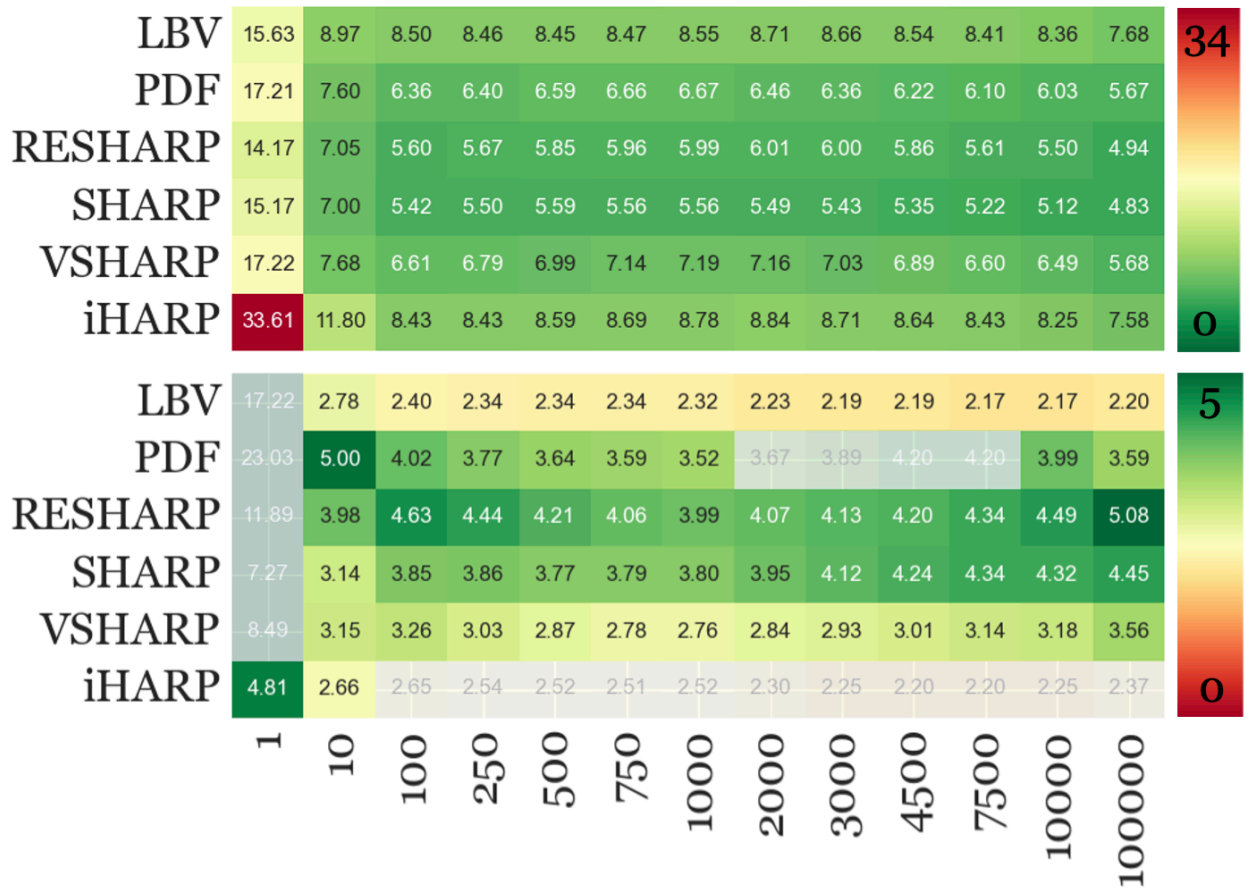

Supplementary Fig. 25 MEDI lambda parameter investigation on the reproducibility (top) and sensitivity (bottom) metric. Y-axis displays the choice of BFR while x-axis shows the lambda (regularization) parameter. Translucent-boxes stand for excluded pipelines due to inconsistent H&S-dependent overtime changes detected in one or more DGM regions.

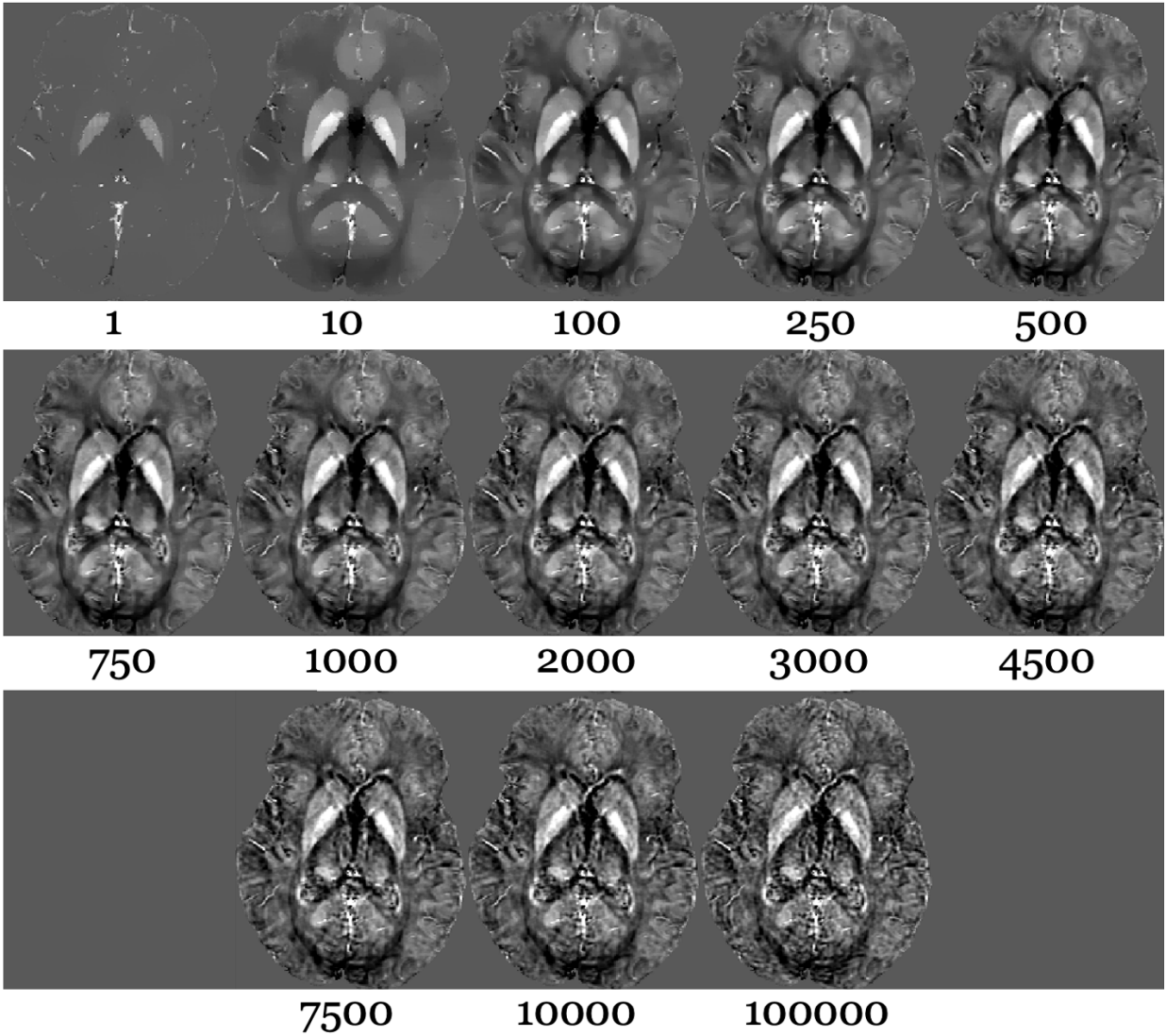

Supplementary Fig. 26 Susceptibility maps from a representative subject in native space using different MEDI lambda parameters with VSHARP as the BFR algorithm. The lambda (λ) parameters are displayed at the bottom of each susceptibility map. Contrast was set manually (-0.08 to 0.15) (ppm).

# Supplementary Tables

Supplementary Table 1. Comments from raters on the visual appearance of the susceptibility maps in the 95$^{th}$ percentile (RESHARP+AMP-PE, HEIDI, and LSQR)

|  | AMP-PE | HEIDI | LSQR |
| --- | --- | --- | --- |
| Rater 1 | Streaking artifacts in the sagittal plane (not as many as LSQR), blurry, pixelated | Sharp, less blurry | Noisy, streaking artifacts in the sagittal plane, inhomogeneous, WM veins visible, not natural. |
| Rater 2 | Reconstruction artifacts | Homogeneous, no visible reconstruction artifacts | Inhomogeneous |
| Rater 3 | Pixelated, too blocky | Homogenous, the best | Streaking artifacts in the coronal plane, not physiological, inconsistent gray matter and WM contrast (inhomogeneous), cloudy, tissue boundaries not sharp |